\documentclass[twocolumn,showpacs,nofootinbib,floatfix,superscriptaddress]{revtex4-1} 
\usepackage{graphicx}
\usepackage{amssymb}
\usepackage{amsmath}
\usepackage{amsthm}
\usepackage{amsfonts}
\usepackage{hyperref}
\usepackage[ruled]{algorithm2e}
\usepackage{diagbox}

\newcommand{\peff}{p_{\mathrm{eff}}}
\newcommand{\pth}{p_{\mathrm{th}}}
\newcommand{\pfail}{p_{\mathrm{fail}}}

\begin{document}

\title{Advantages of versatile neural-network decoding for topological codes}

\date{\today}

\author{Nishad Maskara}
\affiliation{Califonia Institute of Technology, Pasadena, CA 91125, USA}
\author{Aleksander Kubica}
\affiliation{Perimeter Institute for Theoretical Physics, Waterloo, ON N2L 2Y5, Canada}
\affiliation{Institute for Quantum Computing, University of Waterloo, Waterloo, ON N2L 3G1, Canada}
\author{Tomas Jochym-O'Connor}
\affiliation{Walter Burke Institute for Theoretical Physics and Institute for Quantum Information \& Matter, California Institute of Technology, Pasadena, CA 91125, USA}

\begin{abstract}
\vspace*{5pt}
Finding optimal correction of errors in generic stabilizer codes is a computationally hard problem, even for simple noise models.
While this task can be simplified for codes with some structure, such as topological stabilizer codes, developing good and efficient decoders still remains a challenge.
In our work, we systematically study a very versatile class of decoders based on feedforward neural networks.
To demonstrate adaptability, we apply neural decoders to the triangular color and toric codes under various noise models with realistic features, such as spatially-correlated errors.
We report that neural decoders provide significant improvement over leading efficient decoders in terms of the error-correction threshold.
Using neural networks simplifies the process of designing well-performing decoders, and does not require prior knowledge of the underlying noise model.
\end{abstract}

\maketitle

\section{Introduction}

Recent small-scale experiments \cite{barends2014, corcoles2015, kelly2015, nigg2014} have shown an increasing level of control over quantum systems, constituting an important step towards the demonstration of quantum error correction~\cite{Kitaev2002, Nielsen2010}.
In order to scale up quantum devices and maintain their computational power, one needs to protect logical information from unavoidable errors by encoding it into quantum error-correcting codes \cite{Shor1995}.
One of the most successful class of quantum codes, stabilizer codes \cite{Gottesman1996}, allows one to detect errors by measuring stabilizer operators without altering the encoded information.
Subsequently, errors can be corrected by implementing a recovery operation.
A classical algorithm, which allows one to find an appropriate correction from the available classical data, i.e., the $\pm 1$ measurement outcomes of stabilizers for the given code, is called a decoder.

Optimal decoding of generic stabilizer codes is a computationally hard problem, even for simple noise models \cite{Iyer2015}.
If codes have some structure, then the task of decoding becomes more tractable and efficient decoders with good performance may be available.
For example, in the case of topological stabilizer codes \cite{Kitaev2003, Bravyi1998, Bombin2006, Bombin2013book, Haah2011}, whose stabilizer generators are geometrically local, any unsatisfied stabilizer returning $-1$ measurement outcome indicates the presence of errors on some qubits in its neighborhood.
By exploiting this pattern, many decoding schemes have been developed, some of which are based on cellular automata \cite{Harrington2004, Hastings2013, Herold2015, Herold2017, Duivenvoorden2017, Dauphinais2017, Kubicathesis}, the Minimum-Weight Perfect Matching algorithm \cite{Dennis2002, Delfosse2014, Nickerson2017}, tensor networks \cite{Bravyi2014, Darmawan2018}, renormalization group \cite{Duclos-Cianci2013, Duclos-Cianci2013a, Breuckmann2017, Bravyi2011a, Brown2015} or other approaches \cite{Delfosse2017a, Delfosse2017}.

Efficient decoders with good performance are often taylor-made for specific codes and are not easily adaptable to other settings.
For instance, despite a local unitary equivalence of two families of topological codes \cite{Kubica2015}, the color and toric codes, one cannot straightforwardly use toric code decoders in the color code setting; rather, some careful modifications are needed \cite{Delfosse2014, Kubicathesis}.
Moreover, decoding strategies are typically designed and analyzed for simplistic noise models, which may not describe well errors present in the experimental setup.
Importantly, the best approach to scalable quantum devices is still under debate and dominant sources of noise are yet to be thoroughly explored.
Thus, it would be very desirable to develop decoding methods without full characterization of quantum hardware, which are adaptable to various quantum codes and realistic noise models.

\begin{table}[h!]
\centering
\begin{tabular*}
{\columnwidth}{@{\extracolsep{\fill} } l  c c c}
\hline\hline
\multicolumn{4}{c}{threshold of the triangular color code}\\
\hline\hline
\diagbox{noise}{decoder} & neural & projection & optimal\\	
\hline
bit-/phase-flip		&	$\sim 19.0\%$		&	$\sim 16.2\%$		&	$20.6(4)\%$ \cite{Katzgraber2009}	\\
depolarizing		&	$\sim 17.5\%$		&	$\sim 12.6\%$		&	$18.9(3)\%$ \cite{Bombin2012}		\\
NN-depolarizing	&	$\sim 15.0\%$		&	$\sim 13.5\%$		&	?		\\
\hline\hline
&&&\\
\hline\hline
\multicolumn{4}{c}{threshold of the triangular toric code with a twist}\\
\hline\hline
\diagbox{noise}{decoder} & neural & MWPM & optimal\\			
\hline
bit-/phase-flip		&	$\sim19.6\%$	&	$\sim19.2\%$		&	$20.68(4)\%$ \cite{Dennis2002}	\\
depolarizing		&	$\sim17.8\%$		&	$\sim15.3\%$		&	$18.9(3)\%$ \cite{Bombin2012}		\\
NN-depolarizing	&	$\sim16.7\%$		&	$\sim14.2\%$		&	?		\\
\hline\hline
\end{tabular*}
\caption{
The error-correction threshold for neural decoders compared with standard decoding methods based on the Minimum-Weight Perfect Matching algorithm and the projection decoder.
Neural decoders were applied to 2D toric and color codes with the code distance up to $d=11$.
Numerical simulations were performed for various noise models, including the nearest-neighbor spatially-correlated depolarizing noise model, assuming perfect syndrome measurements.
Threshold error rates are expressed in terms of the effective error rate $\peff$; see Section~\ref{sec_noisemodel} for details.
}
\label{tab_thresholds}
\end{table}

The main goal of our work is to systematically explore recently proposed decoding strategies based on artificial neural networks \cite{Torlai2017, Baireuther2017, Krastanov2017, Varsamopoulos2017, Breuckmann2017a}.
We consider two-step decoding.
In step 1, for any given configuration of unsatisfied stabilizers we deterministically find a Pauli operator, which returns corrupted encoded information into the code space.
After this step, all stabilizers are satisfied but a non-trivial logical operator may have been implemented by the attempted Pauli correction combined with the initial error.
In step 2, we use a feedforward neural network to determine what (if any) non-trivial logical operator is likely to be introduced in step 1, so that we can account for it in the recovery. 
We emphasize that step 2 is a classification problem, particularly well-suited for machine learning.

In our work, we convincingly demonstrate the versatility of neural decoders by applying them to two families of codes, the two-dimensional (2D) triangular color and toric codes, under different noise models with realistic features, such as spatially-correlated errors.
We observe that, irrespective of the noise models, neural-network decoding outperforms standard strategies, including the Minimum-Weight Perfect Matching algorithm \cite{Dennis2002} and the projection decoder \cite{Delfosse2014}; see Table~\ref{tab_thresholds}.
It is worth emphasizing that only the training datasets, but not the explicit knowledge of the noise models or the geometric structure of the codes, were needed to train neural decoders.
We also analyze how computational costs of training and neural network parameters scale with the growing code distance.
Our work indicates that due to its adaptability neural-network decoding is a promising error-correction method, which can be used in a wide range of future small-scale quantum devices, especially if the dominant sources of errors are not well characterized.

The organization of the article is as follows.
We start by discussing quantum error correction from the perspective of topological codes, the triangular color code and the toric code with a twist.
In particular, in Section~\ref{sec_decodingreduction} we explain how to construct the excitation graph, which leads to an efficient algorithm for step 1 of the neural decoder.
In Section~\ref{sec_noisemodel} we introduce a new notion of the effective error rate, which allows us to easily compare threshold error rates for different noise models.
Then, we describe neural decoding and its performance under different noise models, including the spatially-correlated depolarizing noise.
In Section~\ref{sec_training} we explain how training of deep neural networks is accomplished by successively increasing the error rate used to generate the training dataset.
This training method likely has significant impact, since it may lead to faster convergence and better final performance of neural networks for quantum error-correction applications.
We conclude the article with the discussion of our results and their implications for future neural decoders used in practice.

\section{Error correction with topological codes}

\subsection{Topological stabilizer codes}

Stabilizer codes \cite{Gottesman1996} are an important class of quantum error-correcting codes \cite{Shor1995} specified by a stabilizer group~$\mathcal{S}$.
The stabilizer group $\mathcal{S}$ is an Abelian subgroup of the Pauli group generated by $n$-qubit Pauli operators $P_1\otimes\ldots \otimes P_n$, where $P_i \in \{I,X,Y,Z\}$ and $-I \not\in\mathcal{S}$.
The logical information is encoded into the codespace, which is the $(+1)$-eigenspace of all the elements of $\mathcal{S}$.
Logical Pauli operators $\overline L \in \mathcal{L}$ are identified with elements of the normalizer $\mathcal{S}$ of the stabilizer group $\mathcal{S}$ in the Paui group.
An operator $L$ which implements a non-trivial logical Pauli operator $\overline L \neq \overline I$ can be chosen to be a product of Pauli operators, which commute with all the elements in the stabilizer group but do not belong to $\mathcal{S}$.
The weight of the minimal-support non-trivial logical Pauli operator determines the distance of the code.

Physical qubits of the stabilizer code can be affected by noise, which can take encoded logical information outside of the codespace.
By measuring stabilizer generators no information about the original encoded state is revealed.
Rather, one effectively projects errors present in the system onto some Pauli operators and subsequently gains some knowledge about them.
The set of unsatisfied stabilizers returning $-1$ measurement outcome is called a syndrome.
The syndrome serves as a classical input to a decoding algorithm, which allows one to find a recovery Pauli operator bringing the corrupted encoded state back to the codespace.
For a special class of stabilizer codes, the CSS codes~\cite{Calderbank1997}, whose stabilizer generators are products of either $X$- or $Z$-type Pauli operators, one can independently correct $Z$- and $X$-type errors using the appropriate $X$- and $Z$-type syndrome.

Topological stabilizer codes \cite{Kitaev2003, Bravyi1998, Bombin2006, Bombin2013book, Haah2011} are a family of stabilizer codes exhibiting particularly good resilience to noise. 
The distinctive feature of topological stabilizer codes is the geometric locality of their generators. 
Namely, physical qubits can be arranged to form a lattice in such a way that every stabilizer generator is supported on a constant number of qubits within some geometrically local region.
At the same time, no logical Pauli operator can be implemented via a unitary acting on physical qubits in any local region.
By enlarging the system size, one increases the distance and error-correction capabilities of the topological code without changing the required complexity of local stabilizer measurements.
This is in stark contrast with other quantum codes, such as concatenated codes \cite{knill1996}, whose stabilizer weight necessarily increases with the distance and thus makes those constructions experimentally more challenging.

\begin{figure}[h!]
(a)\includegraphics[width=.7\columnwidth]{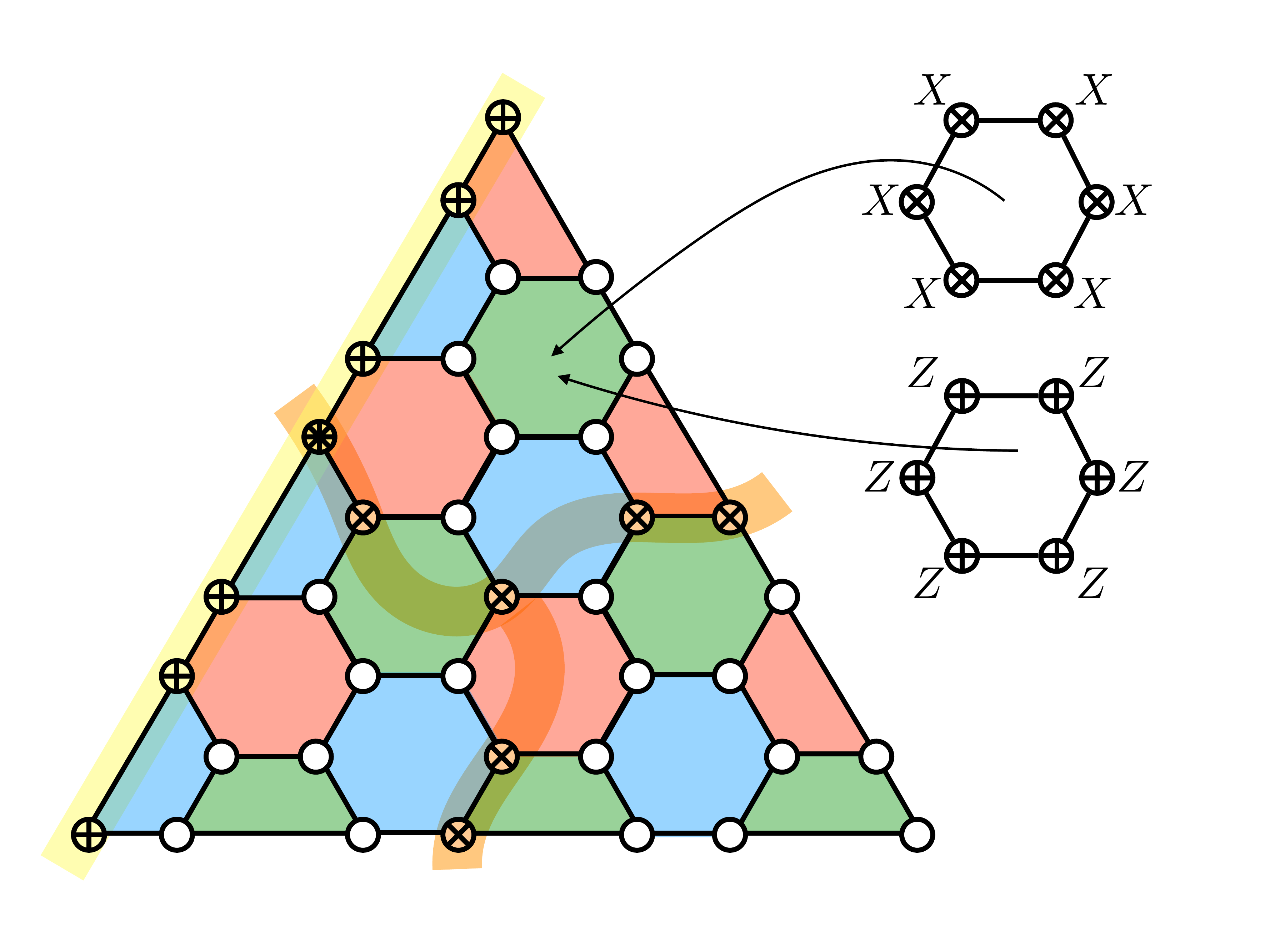}
(b)\includegraphics[width=.7\columnwidth]{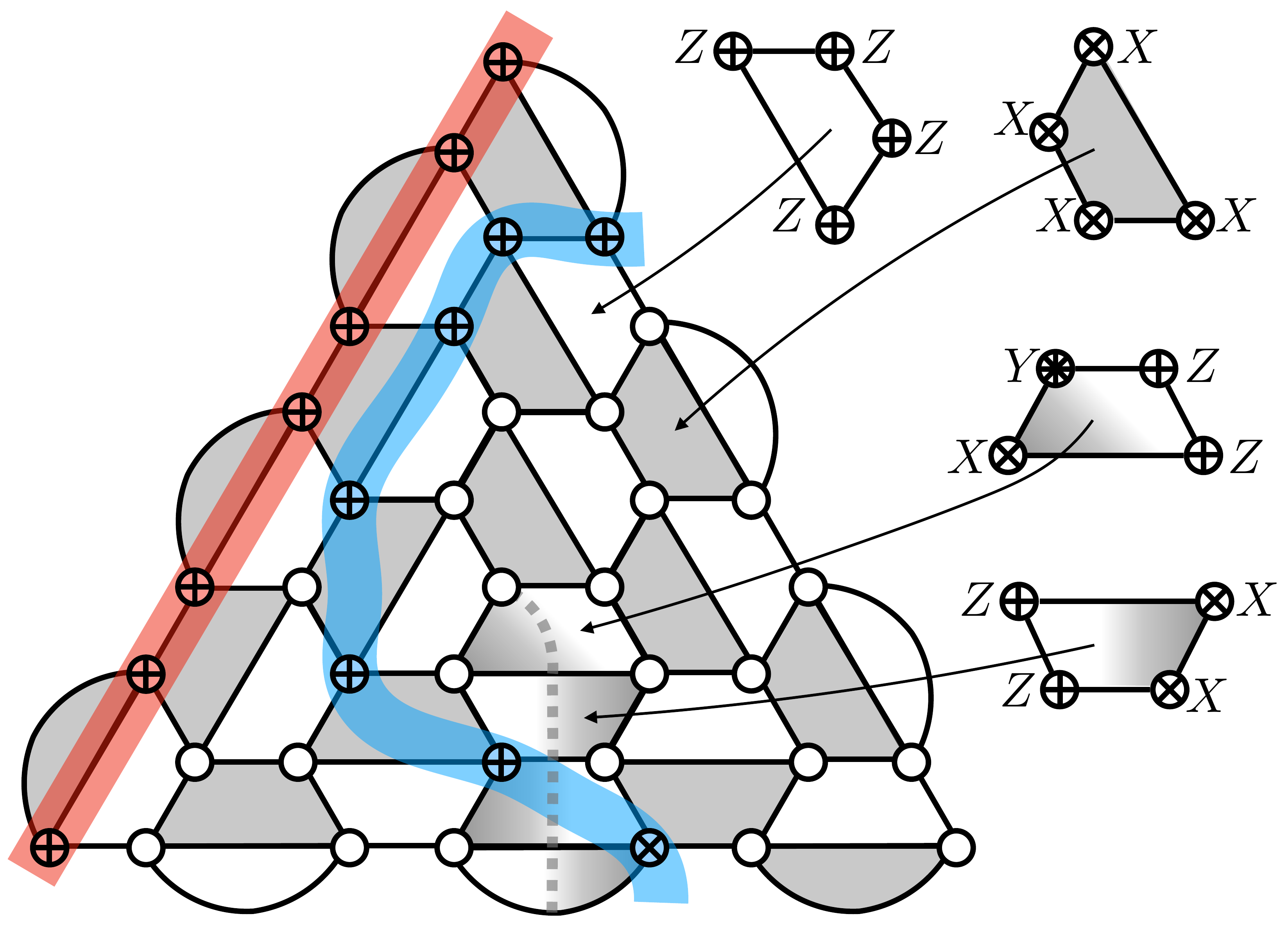}
\caption{
(a) 2D triangular color code on a patch of the hexagonal lattice with $3$-valent vertices and $3$-colorable faces.
Every face supports both $X$- and $Z$-stabilizers.
The string of Pauli $Z$ operators (yellow $\oplus_1$) implements a logical~$\overline Z$ operator, while the string of Pauli $X$ operators (orange $\otimes_2$) implements a logical~$\overline X$.
Both operators connect all three boundaries.
(b) 
2D triangular toric code with a twist.
Dark and white faces support $X$- and $Z$-stabilizers, respectively.
Depending on the coloring of mixed dark/white faces along a 1D defect line (dashed line), stabilizers are mixed products of Pauli $X$ and $Z$.
Red and blue strings depict two equivalent representatives of a logical~$\overline Z$ operator.
Upon crossing the defect line, the string changes from $X$-type (blue $\otimes_1$) to $Z$-type (blue $\oplus_1$).
}
\label{fig_logicalOps} 
\end{figure}

Two well-known examples of topological stabilizer codes are the toric and color codes.
The triangular color code is defined on a two-dimensional lattice with a boundary, whose vertices are 3-valent
\footnote{All the vertices are 3-valent except for three corner vertices on the boundary.}
and faces $f\in F$ are 3-colorable; see Fig.~\ref{fig_logicalOps}(a).
Qubits are identified with vertices.
The color code is a CSS code and its stabilizer group is defined as follows
\begin{equation}
\mathcal{S}_{CC} = \langle X_f,Z_f | f\in F \rangle,
\end{equation}
where $X_f$ and $Z_f$ are Pauli $X$ and $Z$ operators supported on all qubits belonging to a face $f\in F$.
Accordingly, $X$- and $Z$-type errors can be independently corrected using the $Z$- and $X$-type syndrome.

The triangular toric code with a twist \cite{Yoder2017} can be defined for the same arrangement of physical qubits as the triangular color code.
Its lattice can be obtained from the color code lattice by keeping all the vertices, adding extra edges and modifying some faces; see Fig.~\ref{fig_logicalOps}(b).
The resulting lattice is 4-valent
\footnote{All the vertices are 4-valent except for three corner vertices on the boundary and one vertex in the bulk, which corresponds to a twist, i.e., the end of the defect line.}
and the faces are 2-colorable, except for the ``mixed'' faces along a 1D defect line.
The color of the face indicates the type of the stabilizer generator identified with that face.
Namely, dark $f\in F_D$ and white $g\in F_W$ faces support $X$-type $X_f$ and $Z$-type $Z_g$ stabilizers.
Depending on the coloring of mixed faces $h\in F_M$, stabilizers $S_h$  are defined to be mixed products of Pauli $X$ and $Z$ operators.
We emphasize that the choice of mixed stabilizer generators along the defect line is needed for the stabilizers $S_h$ to commute with $X_f$ and $Z_g$ for all $f\in F_D, g\in F_W, h\in F_M$. The full stabilizer group is thus given by
\begin{equation}
\mathcal{S}_{TC} = \langle X_f,Z_g, S_h | f\in F_D, g\in F_W, h\in F_M\rangle.
\end{equation}
We remark that due to mixed stabilizer generators it is not possible to decode $X$ and $Z$ errors independently.

Logical Pauli operators of the 2D topological stabilizer codes can be thought of as deformable non-contractible 1D string-like operators.
In the case of the triangular color and toric codes, logical operators connect certain boundaries as depicted in Fig.~\ref{fig_logicalOps}.

\subsection{Quasiparticle excitations}

It is illustrative to establish a connection between quantum error-correcting codes and quantum many-body systems described by commuting Hamiltonians.
For a topological stabilizer code with the stabilizer group $\mathcal{S}$ we can define a commuting \emph{stabilizer Hamiltonian} $H(\mathcal{S})$ to be a sum of stabilizer generators of $\mathcal{S}$ with a negative sign.
In particular, for the color code and the toric code with a twist we choose their stabilizer Hamiltonians to be
\begin{eqnarray}
H_{CC} &=& -\sum_{f\in F}X_f -\sum_{f\in F} Z_f, \\
H_{TC} &=& -\sum_{f \in F_D} X_f -\sum_{g \in F_W}  Z_g -\sum_{h \in F_M}  S_h.
\end{eqnarray}
Note that all the terms in the stabilizer Hamiltonian $H(\mathcal{S})$ are mutually commuting, thus any eigenstate of $H(\mathcal{S})$ has to be an eigenstate of every single term.
Since eigenstates of stabilizer generators can only have $\pm 1$ eigenvalues, we conclude that the code space defined as the $(+1)$-eigenspace of all the elements of $\mathcal{S}$ coincides with the ground space of $H(\mathcal{S})$.

We can think of errors affecting information encoded in the topological stabilizer code as operators creating localized quasiparticle excitations in the related quantum many-body system.
Namely, consider any Pauli error which anticommutes with some stabilizer generators.
The error moves the encoded logical state outside the code space or, equivalently, the ground state outside the ground space.
The resulting state is excited in the sense that its energy is larger than the ground space energy by the amount proportional to the number of violated stabilizer Hamiltonian terms.
The unsatisfied stabilizer terms can be identified with quasiparticle excitations
\cite{Wilczek1982, Kitaev2003, Preskill1999, Bombin2007}.
Depending on whether the unsatisfied stabilizer is of $X$- or $Z$-type, we will call the excitation electric $e_K$ or magnetic $m_K$.
\footnote{For the mixed stabilizers along the defect line, there is ambiguity in associating the type of the excitation since the electric and magnetic excitations are exchanged upon crossing the defect line. 
Thus, we would refer to those excitations without specifying their type.
}
The subscript $K$ indicates the color of the face supporting the excitation.
In particular, for the toric code we can only have $e_D$ and $m_W$, whereas the color code excitations can be supported on faces of any color, i.e., $e_K$ and $m_K$ for any $K\in \{ R, G, B\}$.

\begin{figure}
(a)\includegraphics[width=.7\columnwidth]{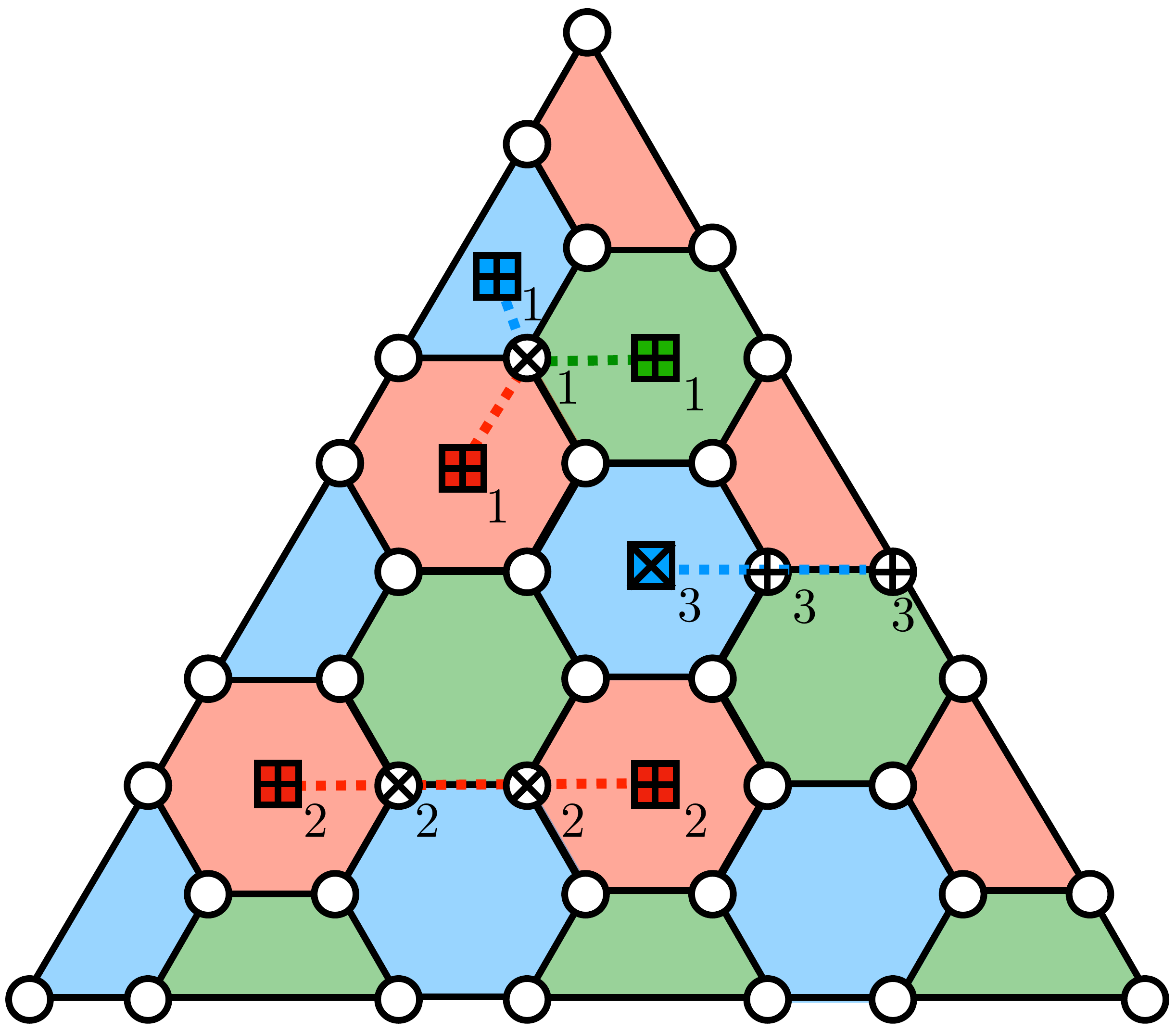}
(b)\includegraphics[width=.7\columnwidth]{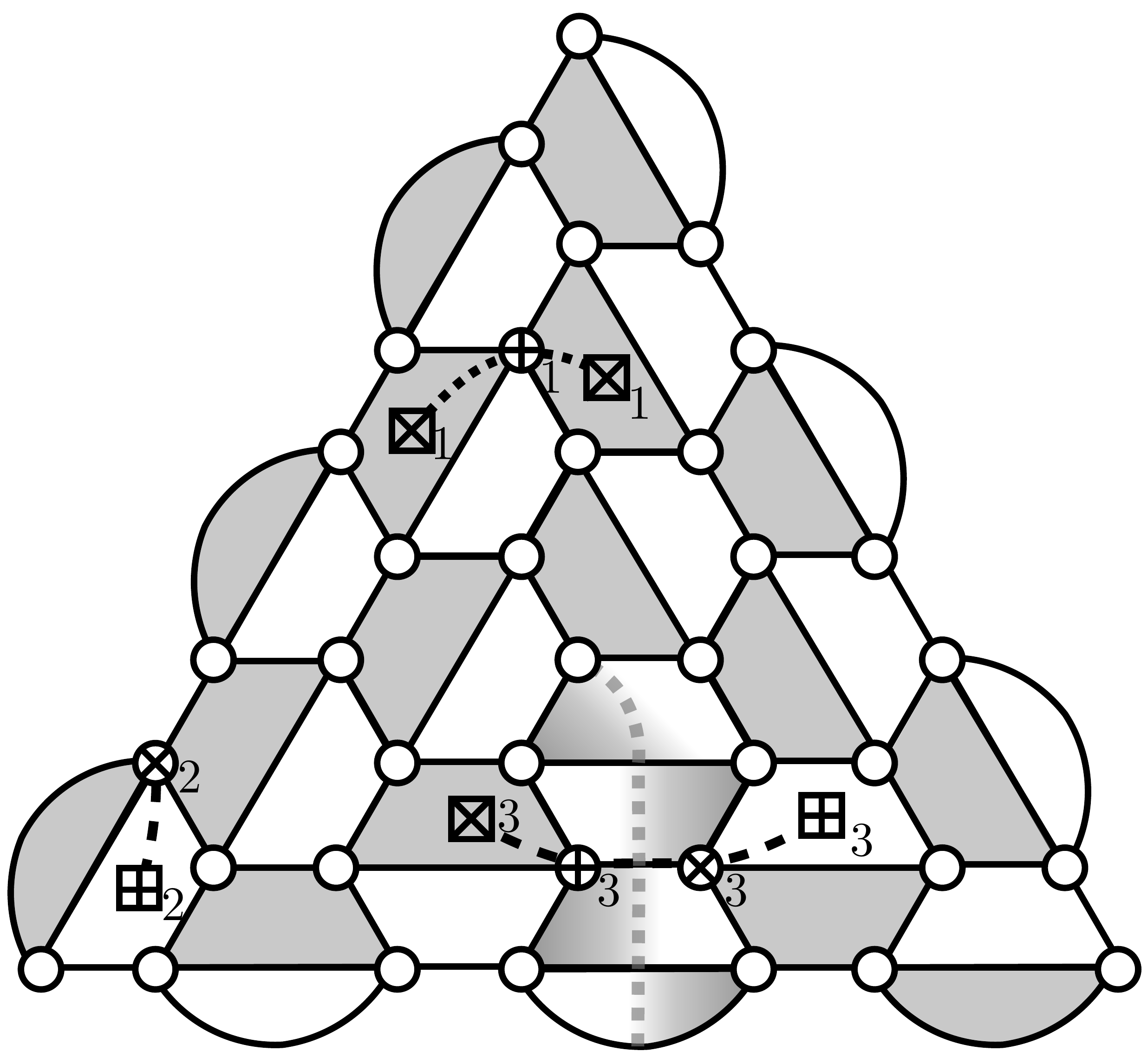}
\caption{
Quasiparticle excitations in the 2D triangular color and toric codes.
(a) A single $X$-error (white $\otimes_1$) in the bulk of the color code leads to three unsatisfied $Z$-stabilizers on neighboring faces, thus creates a triple of magnetic excitations (red, green and blue $\boxplus_1$).
A string of $X$-errors (white $\otimes_2$) creates a pair of magnetic excitations (red $\boxplus_2$).
A string of $Z$-errors (white $\oplus_3$) terminating at the blue boundary creates a single electric excitation (blue $\boxtimes_3$).
(b) A single $Z$-error (white $\oplus_1$) in the bulk of the toric code with a twist leads to two unsatisfied $X$-stabilizers on neighboring dark faces, thus creates a pair of electric excitations (gray $\boxtimes_1$).
A single $X$-error (white $\otimes_2$) on the rough boundary creates a single magnetic excitation (white $\boxplus_2$).
A pair of electric (gray $\boxtimes_3$) and magnetic (white $\boxplus_3$) can be created by a string of errors (white $\otimes_3$ and $\otimes_3$) across the defect line (dashed line).
}
\label{fig_toricandcolor} 
\end{figure}

In order to understand excitation configurations arising from any Pauli errors, it suffices to know what excitations geometrically local Pauli operators can create and how to combine them.
We now discuss these constraints, also known as \emph{fusion rules} for topological stabilizer codes.
In case of the toric code, a single-qubit Pauli $X$ or $Z$ error on the qubit in the bulk of the system violates two $Z$- or $X$-type stabilizers on neighboring faces and thus necessarily creates two excitations of the same type, either magnetic or electric; see Fig.~\ref{fig_toricandcolor}(b).
If two errors with non-overlapping support independently create the same excitation on a face $f\in F$, then the product of both errors will not create any excitation at that location.
For an illustration, let us consider two single-qubit errors $X_i$ and $X_j$ on qubits $i$ and $j$ belonging to the edge  $\{ i,j \}$.
Each error independently creates a magnetic excitation on the face $f$ containing the edge  $\{ i,j \}$; however, the combined error $X_i X_j$ results in no excitation on $f$.  
The above discussion can be summarized by the toric code fusion rules
\begin{equation}
e_D \times e_D = m_W \times m_W = 1,
\label{eq_TCfusion}
\end{equation}
which express the fact that in the bulk excitations of the same type can only be created (by geometrically local operators) or annihilated in pairs.
Note that $1$ denotes no excitation.

The fusion rules for the color code are slightly more complicated than for the toric code. Namely, we have
\begin{eqnarray}
e_K \times e_K = m_K \times m_K &=& 1,\\
\label{eq_fusion1}
e_R \times e_G \times e_B = m_R \times m_G \times m_B &=& 1,
\label{eq_fusion2}
\end{eqnarray}
where $K\in\{R,G,B\}$.
Similarly as for the toric code, combining two excitations of the same type and color results in no excitation.
However, in the bulk of the color code it is also possible to create (by a local operator) or annihilate a \emph{triple} of excitations.
We can see that by considering a single-qubit Pauli $X$ or $Z$ error.
It violates three $Z$- or $X$-type stabilizers on neighboring red, green and blue faces and thus creates a triple of magnetic or electric excitations; see Fig.~\ref{fig_toricandcolor}(a).

The topological stabilizer codes we consider are defined on lattices with boundaries.
By acting with a local Pauli operator on the qubits near the boundary of the system it is possible to create or annihilate a \emph{single} magnetic or electric excitation.
We emphasize that the type of the boundary determines the type of the allowed excitation \cite{Levin2013}.
For the triangular toric code, there are two types of boundaries, rough or smooth \cite{Bravyi1998}, and a single electric (respectively magnetic) excitation can only be created on the rough (smooth) boundary; see Fig.~\ref{fig_toricandcolor}(b).
In case of the triangular color code, there are three types of boundaries, red, green or blue \cite{Bombin2006}, and single electric and magnetic excitations of given color can be created on the boundary of the matching color; see Fig.~\ref{fig_toricandcolor}(a).

Once a quasiparticle excitation is created, it can always be moved in the bulk of the 2D topological stabilizer code by applying an appropriate 1D string-like Pauli operator \cite{Bombin2014}.
Given fusion rules, the excitation movement can be understood as a process of creating pairs of excitations along some path and fusing them together with the initial one, which results in the excitation changing its position.
When the quasiparticle excitation moves its type does not change, unless it passes through a \emph{defect line}.
A defect line, also known as a transparent domain wall\footnote{
A transparent domain wall can be thought of as an automorphism of the excitation labels which preserves the braiding and fusion rules of the quasiparticle excitations.}
\cite{Bombin2010, Bombin2011, Kitaev2012}, is a 1D object, along which the stabilizer generators are appropriately modified.
In case of the triangular toric code with a twist, one chooses stabilizers on faces intersected by the defect line to be mixed products of Pauli $X$ and $Z$ operators; see Fig.~\ref{fig_logicalOps}(b).
When an electric excitation $e_D$ crosses the defect line, it becomes a magnetic excitation $m_W$, and vice versa, $e_D \leftrightarrow m_W$.
We emphasize that logical Pauli operators for the triangular color and toric codes can be implemented by creating a single excitation on one of the boundaries and transporting it to the other boundary, where it can annihilate; see Fig.~\ref{fig_logicalOps} for examples of logical operators.

We remark that there are only two possible types of defect lines in the toric code, one of which is trivial.
However, in case of the color code, there are 72 different defect lines \cite{Yoshida2015}.
We encourage readers to explore \cite{Kesselring2018} for an illuminating discussion of all the possible boundaries and defect lines in the 2D color code.

\subsection{Decoding of topological codes as a classification problem}
\label{sec_decodingreduction}

\begin{figure}[h!]
(a)\includegraphics[width=.8\columnwidth]{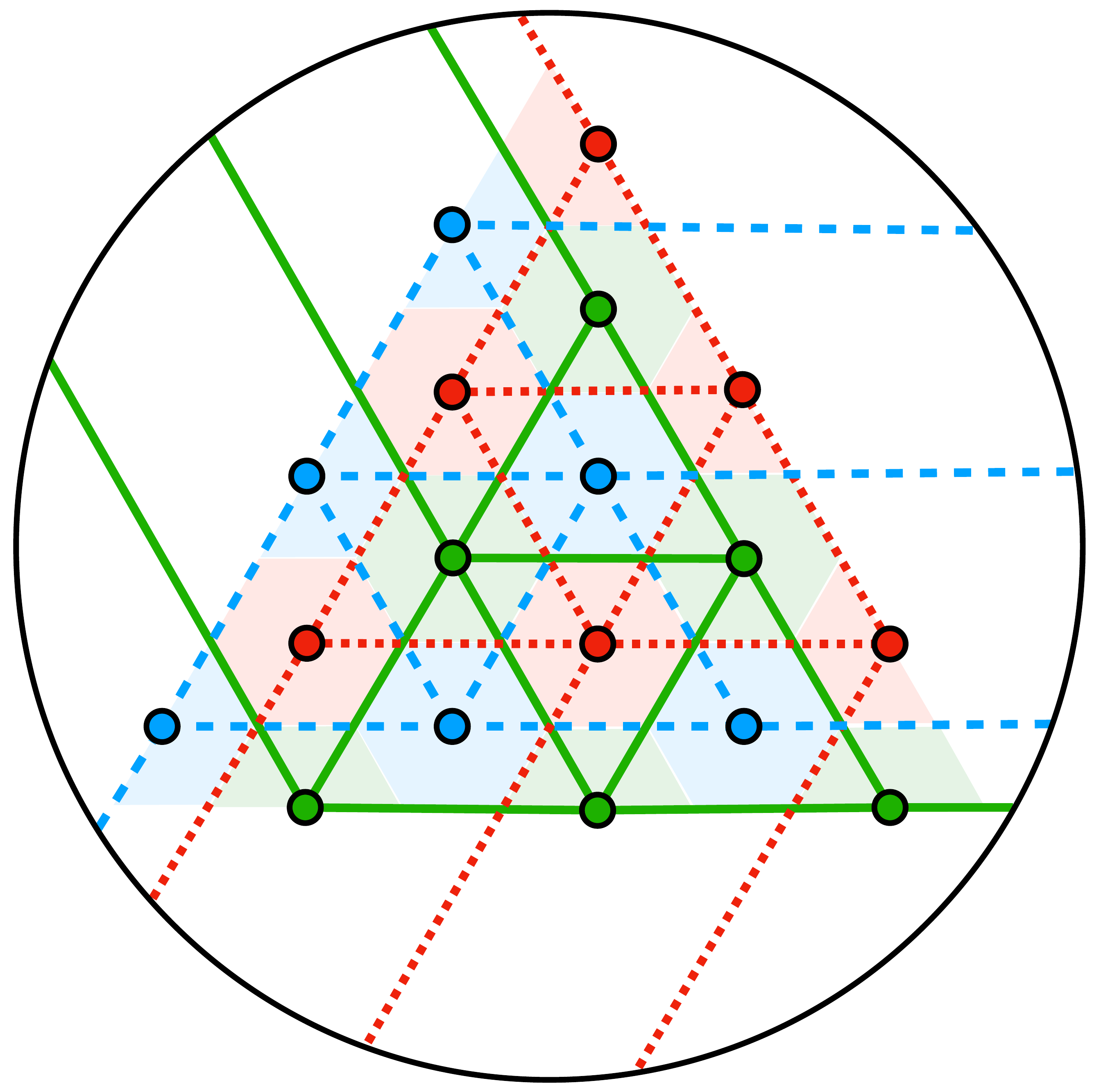}
(b)\includegraphics[width=.8\columnwidth]{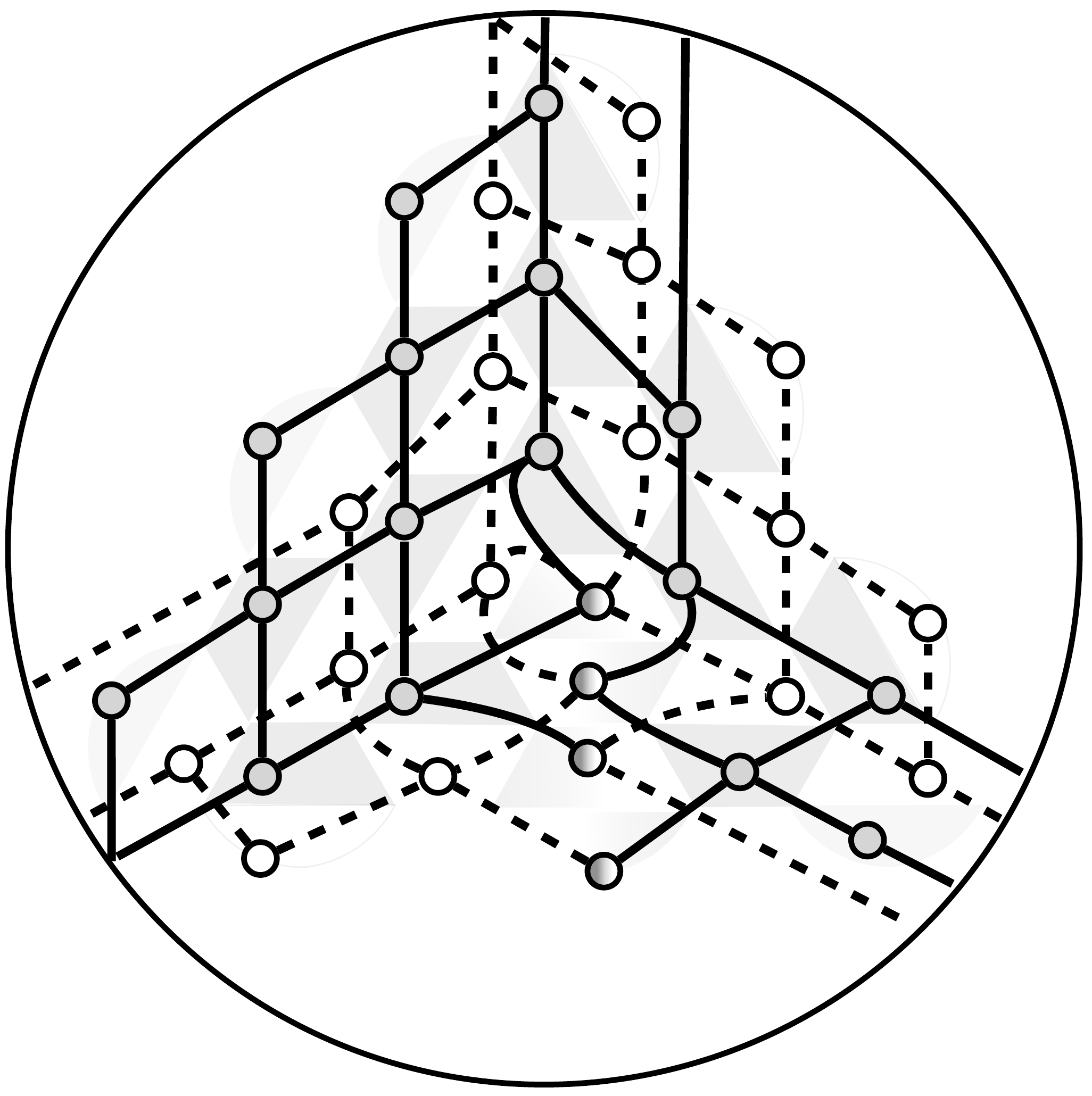}
\caption{
Construction of the excitation graph $G = (V,E)$ for (a) the color code and (b) the toric code with a twist.
For every face $f$ of the topological code lattice we add a vertex $v_f$ to the set of vertices $V$ of $G$.
We also include the boundary vertex $w$ (enclosing circle) in $V$.
(a) It is not possible to move a single excitation in the bulk (without creating more excitations) by applying a single-qubit operator.
However, since a two-qubit operator $XX$ or $ZZ$ can move an excitation between two nearby faces $f$ and $g$ of the same color, we add an edge $\{ v_f,v_g \}$ to $E$.
(b) A single-qubit Pauli $X$ or $Z$ error can move an excitation between two neighboring faces $f$ and $g$ of the same color, thus we add an edge $\{ v_f,v_g \}$ between $v_f$ and $v_g$ to the set of edges $E$ of $G$.
We connect a vertex $v_f$ with the boundary vertex $w$ if one can create a single excitation on $f$ by (a) a single- or two-qubit operators and  (b) a single-qubit operator.
Note that in (a) we depict only a part of the excitation graph corresponding to electric excitations and $Z$-type errors, since the part for magnetic excitations and $X$-type errors is identical.
}
\label{fig_hypergraph} 
\end{figure}

As we already discussed, generic errors affect the encoded information by moving it outside the code space, which results in some stabilizers being unsatisfied.
A classical algorithm which takes the syndrome as an input and finds an appropriate recovery restoring all stabilizers to yield $+1$ measurement outcome is called a \emph{decoder}.
For stabilizer codes the recovery operator is a Pauli operator.
We say that decoding is successful if no non-trivial logical operator has been implemented by the recovery combined with the error.

We can view decoding as a process of removing quasiparticle excitations from the system and returning the state to the ground space of the stabilizer Hamiltonian.
To facilitate the discussion, we introduce an \emph{excitation graph} $G = (V,E)$, which captures how the excitations can be moved (and eventually removed) within the lattice of the topological stabilizer code.
The vertices $V$ of the excitation graph $G$ correspond to the (possible locations of) quasiparticle excitations.
Note that there is one vertex for every single electric, as well as for magnetic excitation.
We also include in $V$ one special vertex $w$, called the boundary vertex.
Two different vertices $v_1,v_2\in V \setminus \{ w \}$ are connected by an edge $\{v_1,v_2\}\in E$ if there is a Pauli operator $P_{v_1, v_2}$ with geometrically local support which can move an excitation from $v_1$ to $v_2$ without creating any other excitations.
We say that $v\in V\setminus \{ w \}$ and the boundary vertex $w$ are connected by an edge $\{v,w\}$ if one can locally create a single excitation at $v$.
In case of the toric and color codes, we restrict our attention to local operators, which are supported on respectively one or at most two neighboring qubits.
We identify the edges $\{v_1, v_2\}$ in $E$ with the local operators $P_{v_1,v_2}$.
We illustrate how to construct the excitation graph in Fig.~\ref{fig_hypergraph}.

We consider a very simple deterministic procedure, the excitation removal algorithm, which efficiently eliminates quasiparticle excitations from the toric and color codes.
Let $Q$ be some Pauli error operator, which results in the excitation configuration $U \subset V \setminus \{ w \}$ in the system.
The input of the algorithm is $U$, but not $Q$.
For every excitation $u\in U$ we find the shortest path $(v_1, v_2, \ldots, v_n)$ in the excitation graph $G$ between $u = v_1$ and the boundary vertex $w = v_n$, where $v_i\in V$ and $\{ v_i,v_{i+1}\} \in E$.
We define an operator $P_{u}$ to be a product of local Pauli operators $P_{v_i,v_{i+1}}$ identified with the edges $\{ v_i,v_{i+1}\}$ along the path $(v_1, v_2, \ldots, v_n)$, namely $P_u = \prod_{i=1}^{n-1}P_{v_i,v_{i+1}}$.
The operator $P_u$ moves an excitation from $u$ to the boundary where it is annihilated.
As the output of the algorithm we choose an operator $R_U = \prod_{u\in U} P_u$. 
We remark that the operator $R_U$ returns the state to the ground space since it removes all the excitations, and thus $R_U Q \in \mathcal{L}$.
At the same time, the output $R_U$ combined with the initial error $Q$ likely implements some non-trivial logical operator.
Thus, the excitation removal algorithm viewed as a decoder would perform rather poorly.

\begin{algorithm}
\caption{excitation removal}
\SetKwInOut{Require}{Require}
\vspace*{2pt}
\Require{the excitation graph $G = (V,E)$}
\vspace*{2pt}
\KwIn{positions $U\subset V \setminus \{ w\} $ of excitations}
\vspace*{2pt}
\KwOut{Pauli operator $R_U$ removing all excitations}
\vspace*{2pt} 
initialize $R_U \leftarrow I$\\
\vspace*{2pt}
for every $u\in U$:\\
\begin{enumerate}
\item find the shortest path $(v_1, \ldots, v_n)$ in $G$ between $u = v_1$ and the boundary vertex $w = v_n$\\
\item find an operator $P_u = P_{v_1,v_2} \cdot\ldots\cdot P_{v_{n-1},v_n}$ corresponding to the path $(v_1, \ldots, v_n)$\\
\item $R_U \leftarrow R_U P_u$\\
\end{enumerate}
\vspace*{-2pt}
\KwRet{$R_U$}
\vspace*{2pt}
\end{algorithm}

Now we explain how to reduce the decoding problem to a classification problem by using the excitation removal algorithm.
The task of classification is to assign labels, typically from some small set, to the elements of some high-dimensional dataset.
In the decoding problem, we know positions $U \subset V \setminus \{ w \}$ of the excitations and  want to find a recovery operator removing all the excitations and implementing the trivial logical operator.
We do not know, however, the Pauli operator $Q$ resulting in the excitation configuration $U$.
Using the excitation removal algorithm we easily find the operator $R_U$.
Clearly, we would be able to successfully decode if we chose $R_U L$ as a recovery operator, where $L$ is any operator implementing the same logical operator $\overline L\in\mathcal{L}$ as $R_U Q$.
Unfortunately, there are many different error operators creating the same configuration of excitations $U$.
We can split all those error operators $Q$ into equivalence classes identified with different logical operators $\overline L$ implemented by $R_U Q$.
Then, for any given excitation configuration $U$ we can find the most probable equivalence class of errors creating $U$.
What we would like to achieve is to label $U$ by a logical operator $\overline L$, which is implemented by the output $R_U$ of the excitation removal algorithm and any operator $Q$ from the most probable class of errors.
Such a problem is well-suited for machine learning techniques, in particular for artificial neural networks.
We defer further discussion of the classification problem to Section~\ref{sec_neuraldecoder}, where we explain it in the context of neural-network decoding.

\subsection{Noise models and thresholds}
\label{sec_noisemodel}

In order to test versatility of neural decoders, we numerically simulate their performance for various noise models.
In particular, we consider the following three Pauli error models specified by just one parameter, the error rate $p$.
\begin{itemize}
\item \emph{Bit-/phase-flip noise}: every qubit is independently affected by an $X$ error with probability $p$, and by a $Z$ error with the same probability $p$.
\item \emph{Depolarizing noise}: every qubit is independently affected with probability $p$ by an error, which is uniformly chosen from three errors $X$, $Y$ and $Z$.
\item \emph{NN-depolarizing noise}: the spatially-correlated depolarizing noise on nearest-neighbor qubits, i.e., every pair of qubits $i$ and $j$ sharing an edge in the lattice is independently affected with probability $p$ by a non-trivial error, which is uniformly chosen from 15 errors of the form $P_i P_j$, where $P_i,P_j \in \{ I,X,Y,Z\}$ and $P_i P_j\neq II$.
\end{itemize}
We emphasize that one should not necessarily think of the aforementioned noise models as accurately describing errors in the experimental setup.
Rather, we choose those models since they are easy to specify and simulate but, at the same time, they also capture realistic noise features, such as spatial correlations of errors, which any good decoder should be able to handle \cite{Nickerson2017}.
In addition, in the current proposed circuit-based models for syndrome measurement~\cite{Fowler2012} correlated errors across neighboring qubits would naturally arise.

We would like to easily compare the bit-/phase-flip, depolarizing and NN-depolarizing noise models.
However, the error rate $p$ has a different meaning depending on the considered model.
This motivates us to introduce a new figure of merit for Pauli error models, the \emph{effective error rate} $\peff$.
For any physical qubit we define the effective error rate $\peff$ to be the probability of any non-trivial error affecting that qubit.
Note that in the scenarios we consider the effective error rate is the same for all the qubits (except for the ones identified with the corner vertices and the twist for the NN-depolarizing noise).
Thus, we can unambiguously talk about the effective error rate without specifying which qubit we are referring to.
For the depolarizing noise we simply have $\peff = p$, whereas for the the bit-/phase-flip noise we find $\peff = 1- (1-p)(1-p) = 2p-p^2$.
In case of the NN-depolarizing noise, the effective error rate depends on the local structure of the lattice.
Namely, if $n$ denotes the number of nearest neighbors for some qubit, then the effective error rate $\peff^{(n)}$ for that qubit can be recursively calculated as
\begin{eqnarray}
\peff^{(n)} &=& \peff^{(n-1)}\left(1-\frac{4}{15}p\right) + \left(1- \peff^{(n-1)} \right)\frac{12}{15}p\\
&=& \frac{4}{5} np + o(p^2),
\end{eqnarray}
where we use $\peff^{(0)} = 0$ and denote by $o(p^2)$ the second-order corrections in $p$. 
In particular, for the analyzed color and toric code lattices we respectively have $\peff^{(3)}$ and $\peff^{(4)}$.

In order to assess the performance of a decoder for the given family of codes with growing code distance $d$ and specified noise model, we use the quantity called the \emph{error-correction threshold}.
The error-correction threshold is defined as the largest $\pth$, such that for all effective error rates $\peff < \pth$ the probability of unsuccessful decoding $\pfail(\peff, d)$ for the code with distance $d$ goes to zero in the limit of infinite code distance,
$\lim_{d\rightarrow \infty} \pfail(\peff, d) = 0$.
Note that in the definition of the threshold we assume perfect stabilizer measurements.
We remark that one typically estimates the threshold $\pth$ by plotting the decoder failure probability $\pfail(\peff, d)$ as a function of the effective error rate $\peff$ for different code distances $d$ and identifying their crossing point; see Figs.~\ref{fig_thresholds_CC}~and~\ref{fig_thresholds_TC}.

\section{Performance of neural-network decoding}
\label{sec_results}

\subsection{Neural decoders}
\label{sec_neuraldecoder}

\begin{figure}[h!]
\includegraphics[width=.95\columnwidth]{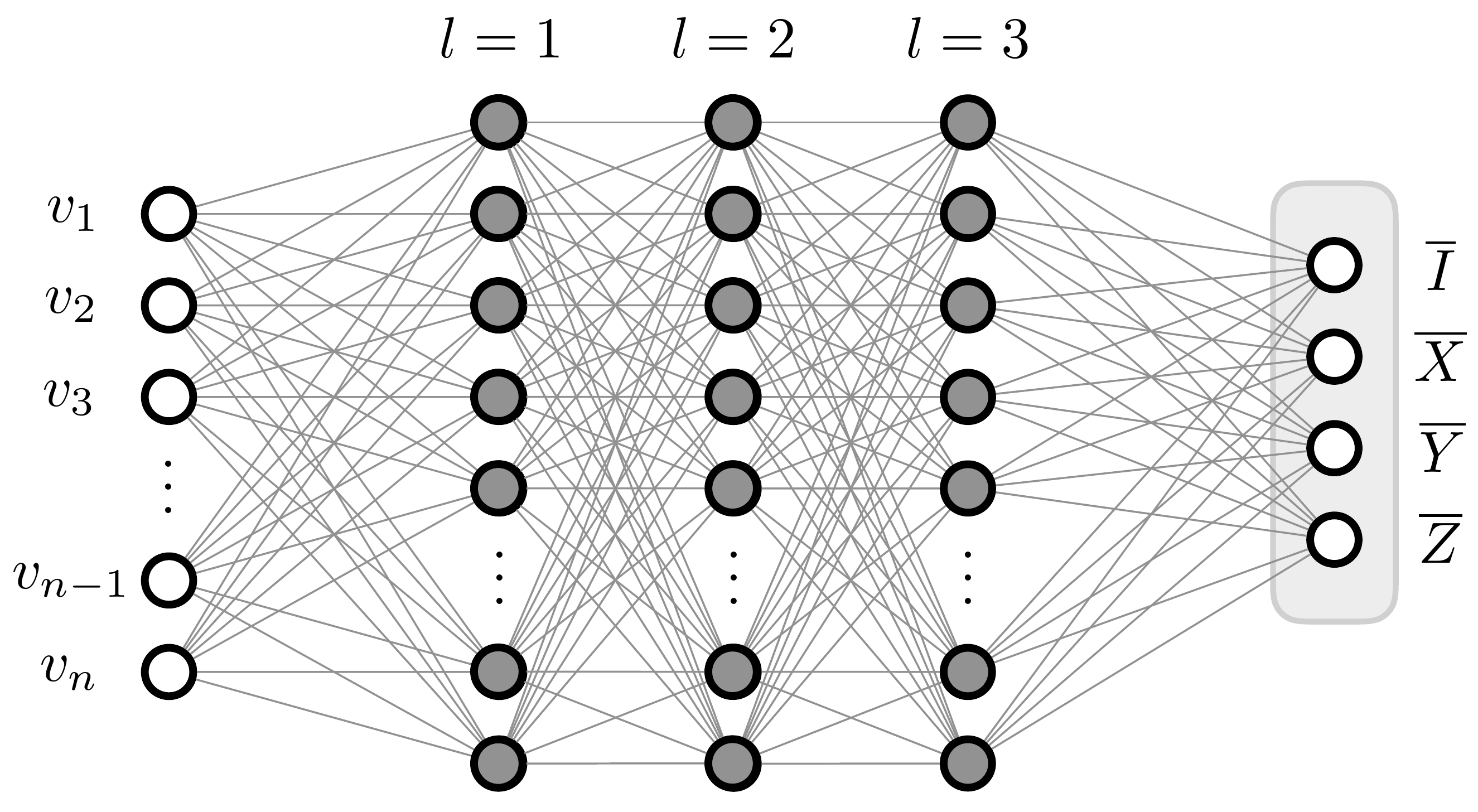}
\caption{
A feedforward neural network with $H_d = 3$ hidden layers.
Each hidden layer has the same number of nodes $N_d$.
Nodes from layer $l+1$ are fully connected with nodes from the preceding layer $l$.
The input layer encodes all the initial excitation configuration $U \subset V \setminus \{ w\}$.
The output layer encodes the likelihood of each logical operator $\overline L \in \{ \overline I, \overline X, \overline Y, \overline Z \}$ assigned to the input configuration $U$.
}
\label{fig_neuralnetwork} 
\end{figure}

We have already seen in Section~\ref{sec_decodingreduction} that the task of successful decoding can be deterministically reduced to the following problem:
for any configuration of excitations $U \subset V \setminus \{ w\}$ created by some unknown Pauli operator $Q$ assign a label $\overline{L}$ from the set of logical operators $\mathcal{L}$, such that $\overline L$ is the logical operator implemented by 
$R_U Q$, where $R_U$ is the output of the excitation removal algorithm with $U$ as the input.
We approach this classification problem by using one of the leading machine learning techniques, feedforward neural networks.
For each code of distance $d$, we train a neural network consisting of $H_d+2$ layers; see Fig.~\ref{fig_neuralnetwork}.
The input layer encodes the configuration of excitations $U$.
Then, there are $H_d$ hidden layers, each containing $N_d$ nodes.
Nodes from layer $l+1$ are fully connected with nodes from the preceding layer $l$.
Every node $\nu$ in layer $l+1$ evaluates an activation function $\sigma (w_\nu \cdot o_{l}  + b_\nu)$ on the output $o_l$ of nodes from layer $l$, where $w_\nu$ and $b_\nu$ are the weights and biases associated with the node $\nu$.
We choose the rectified linear unit activation function $\sigma (x) = \max (0,x)$.
The output layer uses the softmax classifier, which converts an output vector to a discrete probability distribution describing the likelihood of different logical operators $\overline{L}\in\mathcal{L}$ being implemented by $R_U Q$.

We are now ready to describe neural-network decoding for topological stabilizer codes.
The neural decoder is an algorithm which returns a recovery operator $R$ for any configuration of excitations $U\subset V \setminus \{ w\}$ created by some unknown operator $Q$.
We emphasize that error operators $Q$ are chosen according to some a priori unknown noise model.
The neural decoders we consider consist of the following two steps.
In step 1, we use a simple deterministic procedure, the excitation removal algorithm, to find a Pauli operator $R_U$, which removes quasiparticle 
excitations by moving them to the boundaries of the system, where they disappear.
In step 2, we use a neural network to guess what are the most likely errors $Q$ resulting in $U$ and which logical operator $\overline L$ is subsequently implemented by $R_U Q$.
As the output, the operator $R_U L$ is returned, where $L$ is any operator implementing the logical operator $\overline L$.
We emphasize that the neural decoder always returns a valid recovery operator but decoding succeeds if and only if the neural network correctly identifies the logical operator $\overline L$ implemented by $R_U Q$.
Moreover, determining the output of the trained neural network is efficient since it reduces to matrix multiplication.
We see that in step 1 we implicitly make use of the excitation graph, which contains information about the topological code lattice and the fusion rules.
However, no information about the topological code is required to train the neural network, which is used in step 2.

\begin{algorithm}
\caption{neural decoder}
\SetKwInOut{Require}{Require}
\vspace*{2pt}
\Require{excitation removal algorithm, trained neural network}
\vspace*{2pt}
\KwIn{locations of excitations $U\subset V \setminus \{ w\}$ created by some unknown operator $Q$}
\vspace*{2pt}
\KwOut{recovery operator $R$}
\vspace*{2pt}
using the excitation removal algorithm with $U$ as the input, find an operator $R_U$\\
\vspace*{2pt}
 using the neural network with $U$ as the input, find the logical operator $\overline{L}$ implemented by $R_U Q$\\
 \vspace*{2pt}
$R \leftarrow R_U L$, where $L$ is any operator implementing $\overline L$\\
\vspace*{2pt}
\KwRet{$R$}
\vspace*{2pt}
\end{algorithm}

We emphasize that the details of step 1 in the neural decoder do not matter as long as the returned operator $R_U$ is found in an efficient deterministic way.
We choose the excitation removal algorithm because it is simple and has an intuitive explanation --- it removes all the excitations by moving them to the boundaries of the system.
We point out that we could use a similar version of the neural decoder for other topological codes (or even codes without geometric structure), as long as we knew how to efficiently find the operator $R_U$.
For instance, if we considered the toric or color codes on a torus, with or without boundaries, then we could always find a simple removal procedure which deterministically moves all excitations of the same color to the same location in the bulk or on the boundary, where they are guaranteed to disappear.
Such a procedure can then be used to create the training dataset for the neural network.
We remark that step 1 becomes more challenging for codes without string-like operators, such as the cubic code~\cite{Haah2011}.

\subsection{Training deep neural networks}
\label{sec_training}

Before a neural network can be used for decoding, it needs to be trained.
We do this via supervised learning, where the network is trained on a dataset of preclassified samples.
Sample Pauli errors are generated using Monte Carlo sampling according to the appropriate probability distribution determined by the noise model.
For each generated error configuration $Q$, we determine the corresponding syndrome, i.e., the excitation configuration $U\subset V \setminus \{ w \}$, which is the input to the neural network.
Then, using the excitation removal algorithm, we find the Pauli operator $R_U$, and check what logical operator $\overline{L}$ is implemented by $R_U Q$.
This allows us to label each input excitation configuration $U$ with the corresponding classification label $\overline{L}$ we want the neural network to output.
We remark that the testing samples used to numerically estimate thresholds are created in the same way as the training samples.
 
Training the neural network can now be framed as a minimization problem.
The network parameters, i.e., the weights and biases, are optimized to minimize classification error on the training dataset.
We use the categorical cross entropy cost function $C$ to quantify the error, namely
\begin{equation}
C = \sum_i \vec{y_i}\cdot \log\left(\vec{f}(\vec{x_i})\right) + (\vec{1}-\vec{y_i}) \cdot \log \left(\vec{1} - \vec{f}(\vec{x_i})\right),
\label{eq_cost}
\end{equation}
where $\vec{y_i}$ is the classification bit-string for the input $\vec{x_i}$, $\vec{f}(\vec{x_i})$ is the likelihood vector returned by the neural network, and $\vec{1} = (1,\ldots,1)$.
Importantly, this cost function is differentiable, which allows us to use backpropagation to efficiently compute the gradient of the cost function with respect to network parameters in a single backwards pass of the network.
The minimization is performed using Adam optimization \cite{kingma2014adam}, a highly effective variant of gradient descent, whose learning parameters do not need to be fine-tuned for good performance.
In practice, we find that Adam optimization converges significantly faster than standard gradient descent, with the effects becoming more pronounced for larger networks.

Instead of computing the cost function on the entire training set, which becomes computationally expensive for very large datasets, we use mini-batch optimization.
This is a standard technique, which estimates the cost function on individual batches, i.e., small subsets of the training datasets; see e.g.~\cite{hinton2012}.
We define a training step as one round of backpropagation and a subsequent network parameter update, using the cost function 
$C$ in Eq.~(\ref{eq_cost}) estimated on a single batch.
The batch size controls the accuracy of this estimate and needs to be manually adjusted.

Until recently, training deep neural networks had been next to impossible.
However, innovations by the machine learning community have made it easy to train extremely deep networks.
We too were unable to successfully train networks with more than three hidden layers, until we implemented two of these improvements: He initialization and batch normalization.
He initialization \cite{he2015delving} ensures that learning is efficient for the rectified linear unit activation function, whereas batch normalization \cite{ioffe2015batch} stabilizes the input distribution for each layer.
Batch normalization makes it possible to train deeper networks, as well as improves performance on shallower three-layer networks.

The training set is generated according to the noise model and some chosen error rates.
Once the neural network is trained, it should be able to successfully label syndromes for error configurations generated at various error rates below the threshold.
In particular, any fine-tuning of the network for specific error rates is not desired.
Since the error syndromes for higher error rates are in general more challenging to classify, it would be desirable to train the neural network mainly on configurations corresponding to error rates close to the threshold.
However, during training of the networks for higher-distance codes and correlated noise models the optimization algorithm is very likely to get stuck in local minima if we start training on the high error-rate dataset directly.
This problem is manifested in the network not effectively learning the noise features and the resulting performance showing only small improvements over random guessing.
A solution we propose is to first pre-train the network on a lower error-rate dataset, and only then use the training data corresponding to the near-threshold error rate; changing of the error rate does not have to be very slow.
We believe that this is an important observation for any future implementations of neural networks for decoding quantum error-correcting codes.
We also speculate that a similar strategy might help to speed up training of neural networks for experimental systems.
Namely, we imagine pretraining the neural network for some simple theoretical error models at low error rates, and then using the experimental data for further training.

\begin{table}[h!]
\centering
\begin{tabular*}
{\columnwidth}{@{\extracolsep{\fill} } l  c c c c c}
\hline\hline
\multicolumn{6}{c}{training cost for the triangular color code}\\
\hline\hline
\diagbox{noise}{parameters} & $d$ & $H_d$ & $N_d$ & $B_d$ & $T_d$ \\	
\hline
bit-/phase-flip		&	5	& 3	& 100	& $10^3$	& $3 \times 10^4$	\\
				&	7	& 5	& 200	& $ 5 \times10^3$	& $5 \times 10^4$	\\
				&	9	& 7	& 400	& $10^4$	& $1.1 \times 10^5$	\\
				&	11	& 9	& 800	& $10^4$	& $2.1 \times 10^5$	\\
depolarizing		&	5	& 3	& 200	& $10^4$ & $1.1 \times 10^5$	\\
				&	7	& 5	& 600	& $10^4$	& $3 \times 10^5$	\\
				&	9	& 7	& 1400	& $10^4$	& $4.1 \times 10^5$	\\
NN-depolarizing	&	5	& 3	& 200	& $5 \times 10^3$ & $6 \times 10^4$	\\
				&	7	& 5	& 400	& $10^4$	& $1.1 \times 10^5$	\\
				&	9	& 7	& 800	& $10^4$	& $2.1 \times 10^5$	\\
				&	11	& 9	& 1600	& $10^4$	& $4.1 \times 10^5$	\\
\hline\hline
&&&\\
\hline\hline
\multicolumn{6}{c}{training cost for the triangular toric code with a twist}\\
\hline\hline
\diagbox{noise}{parameters} & $d$ & $H_d$ & $N_d$ & $B_d$ & $T_d$ \\	
\hline
bit-/phase-flip		&	5	& 3	& 100	& $10^3$	& $3 \times 10^4$	\\
				&	7	& 5	& 200	& $10^4$	& $6 \times 10^4$	\\
				&	9	& 7	& 400	& $10^4$	& $1.6 \times 10^5$	\\
				&	11	& 9	& 800	& $10^4$	& $2.6 \times 10^5$	\\
depolarizing		&	5	& 3	& 200	& $5 \times 10^3$ & $3 \times 10^4$	\\
				&	7	& 5	& 600	& $10^4$	& $1.1 \times 10^5$	\\
				&	9	& 7	& 1200	& $10^4$	& $2.1 \times 10^5$\\
NN-depolarizing	&	5	& 3	& 200	& $5 \times10^3$ & $6 \times 10^4$	\\
				&	7	& 5	& 400	& $10^4$	& $1.1 \times 10^5$	\\
				&	9	& 7	& 800	& $10^4$	& $2.1 \times 10^5$	\\
				&	11	& 9	& 1600	& $10^4$	& $4.1 \times 10^5$\\
\hline\hline
\end{tabular*}
\caption{
Optimal neural-network hyperparameters of the neural decoder for the triangular color code (top) and the triangular toric code with a twist (bottom) with distance $d$ under different noise models.
Hyperparameters varied are: the number of hidden layers $H_d$, the number of nodes in the hidden layer $N_d$, the batch size $B_d$ and the number of training steps~$T_d$.
The total number of training samples seen during training is $B_d T_d$.
}
\label{tab_hyperparameters}
\end{table}

\subsection{Selecting neural-network hyperparameters}

In addition to network parameters, there are also hyperparameters which cannot be trained via backpropagation.
These include the number of hidden layers $H_d$, the number of nodes per hidden layer $N_d$, the size of each batch $B_d$, and the total number of training steps $T_d$.
We optimize these parameters using a grid search based approach; see Table~\ref{tab_hyperparameters} for the optimal values we find.
A heuristic rule for determining the size of a well-performing neural network for the code with distance $d$ is to use $H_d = d-2$ hidden layers and $N_d \propto 2^{d/2}$ nodes per layer.
Whether or not this exponential trend continues for larger code sizes is an open question.

We notice that very large training sets are needed for optimal performance.
In order to save on computational memory, we choose to generate training samples in parallel to training, since it can be done efficiently.
Note that with this strategy the number of different samples seen during training is $B_d T_d$.
We observe that the training time appears to scale exponentially with code distance, approximately doubling as the distance increases by two.

We find evidence that there is some minimal batch size below which the gradient estimates are too noisy for the network to converge to a solution that outperforms random guessing.
However, increasing the batch size beyond that minimal value does not improve the final network performance.
Rather, it reduces the number of training steps needed for convergence, but with diminishing returns.
The batch size we choose is primarily optimized to minimize the training time.

\subsection{Thresholds of neural decoders}

\begin{figure*}[ht!]
\centering
(a)\includegraphics[width= 0.29\textwidth]{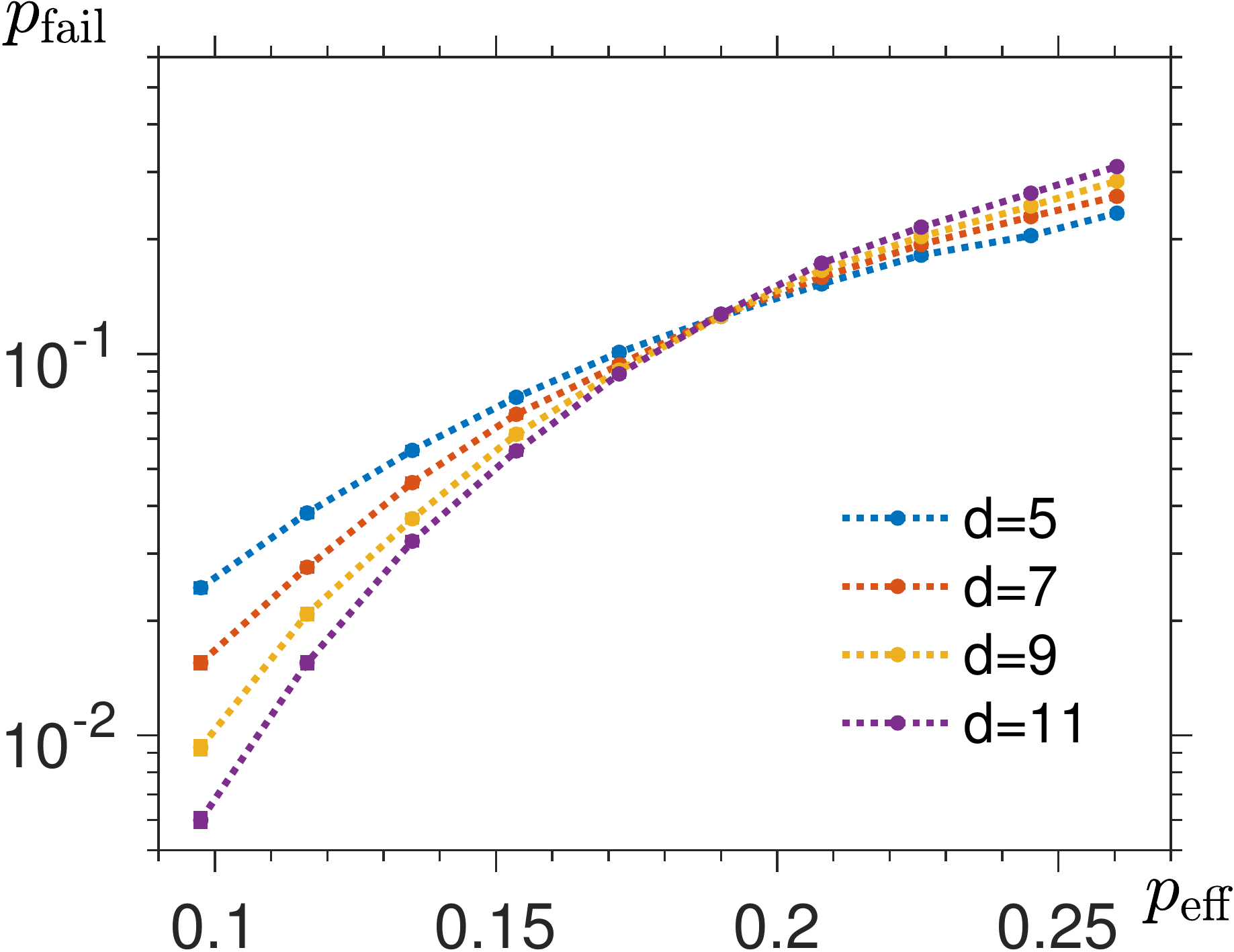}\quad
(b)\includegraphics[width= 0.29\textwidth]{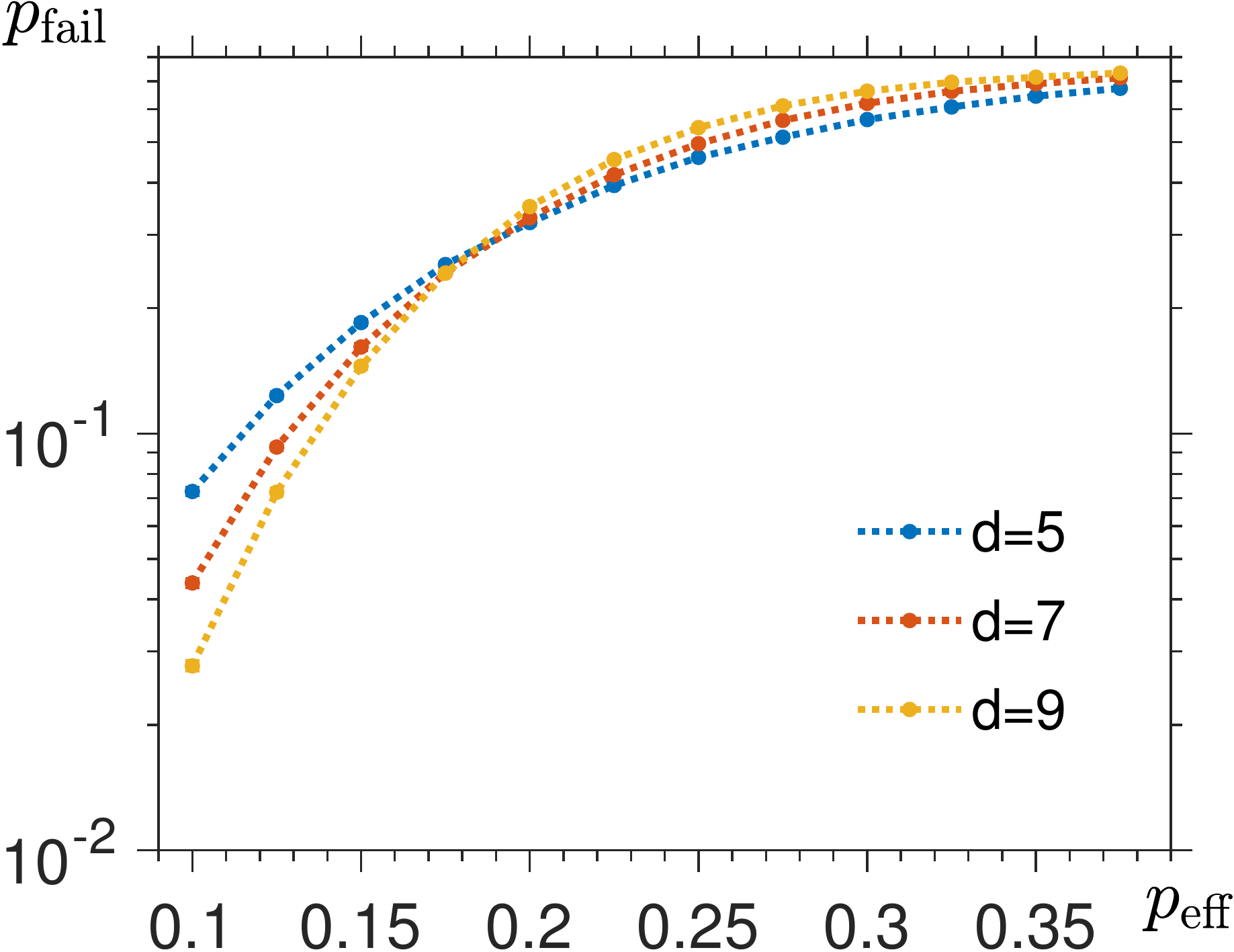}\quad
(c)\includegraphics[width= 0.29\textwidth]{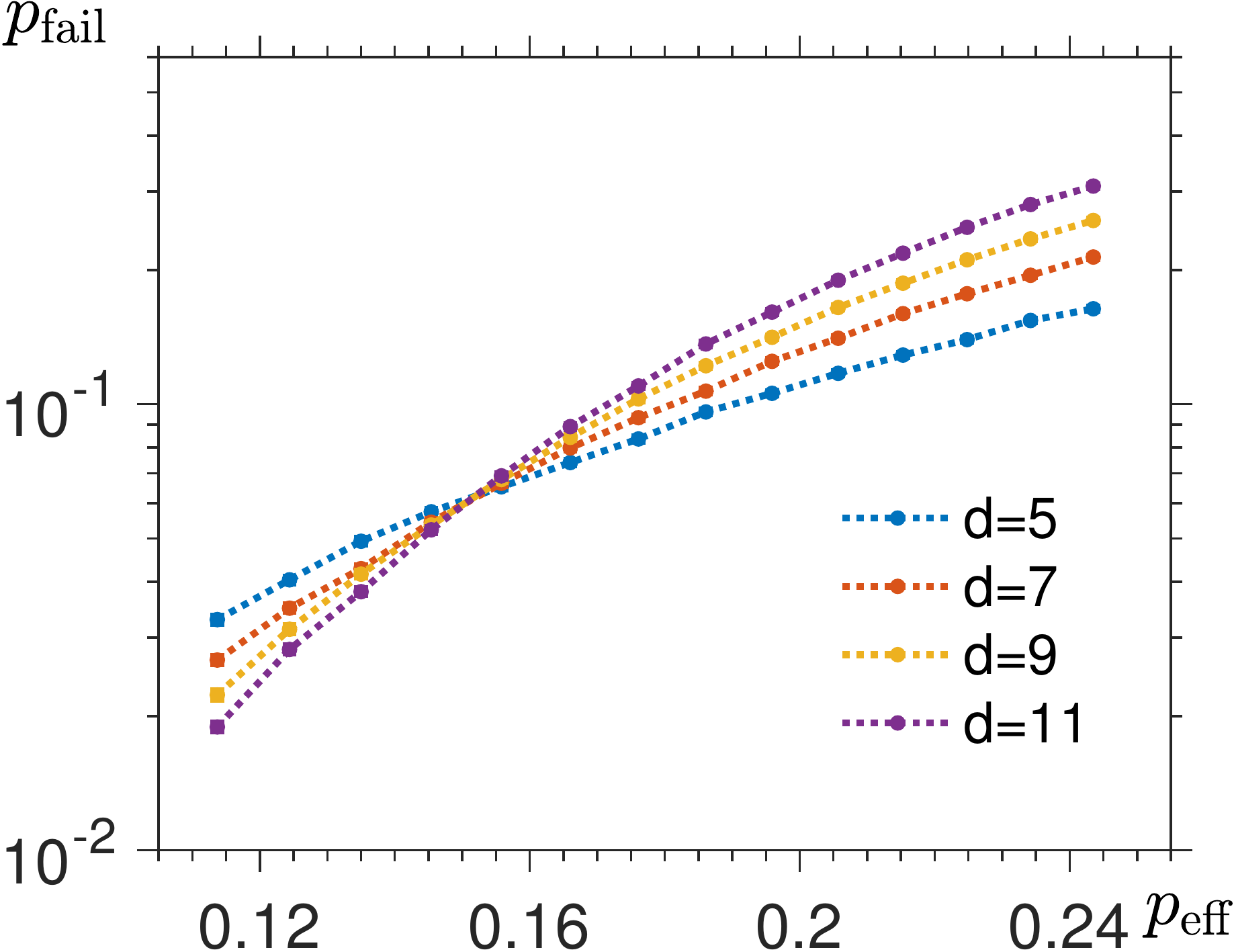}
(d)\includegraphics[width= 0.29\textwidth]{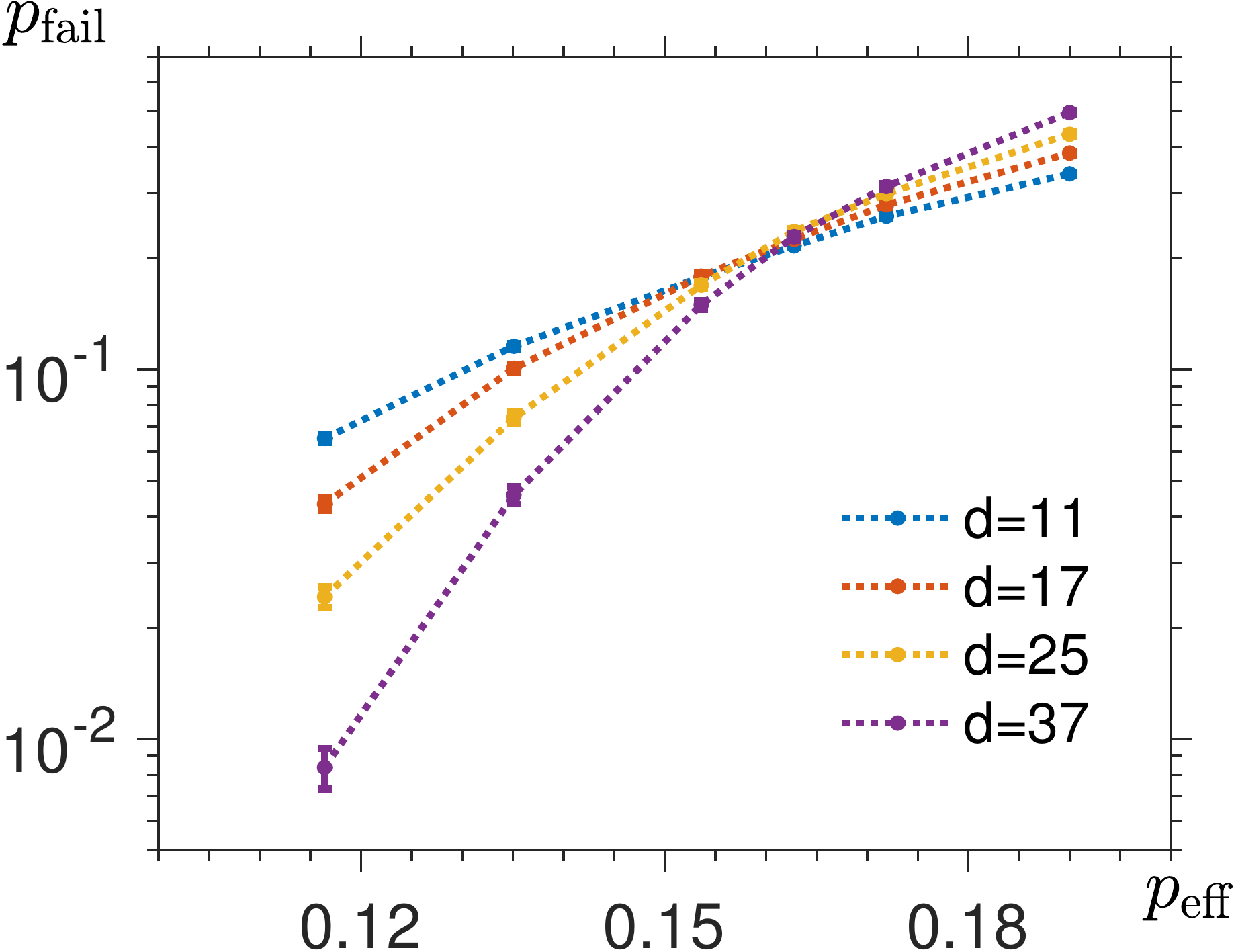}\quad
(e)\includegraphics[width= 0.29\textwidth]{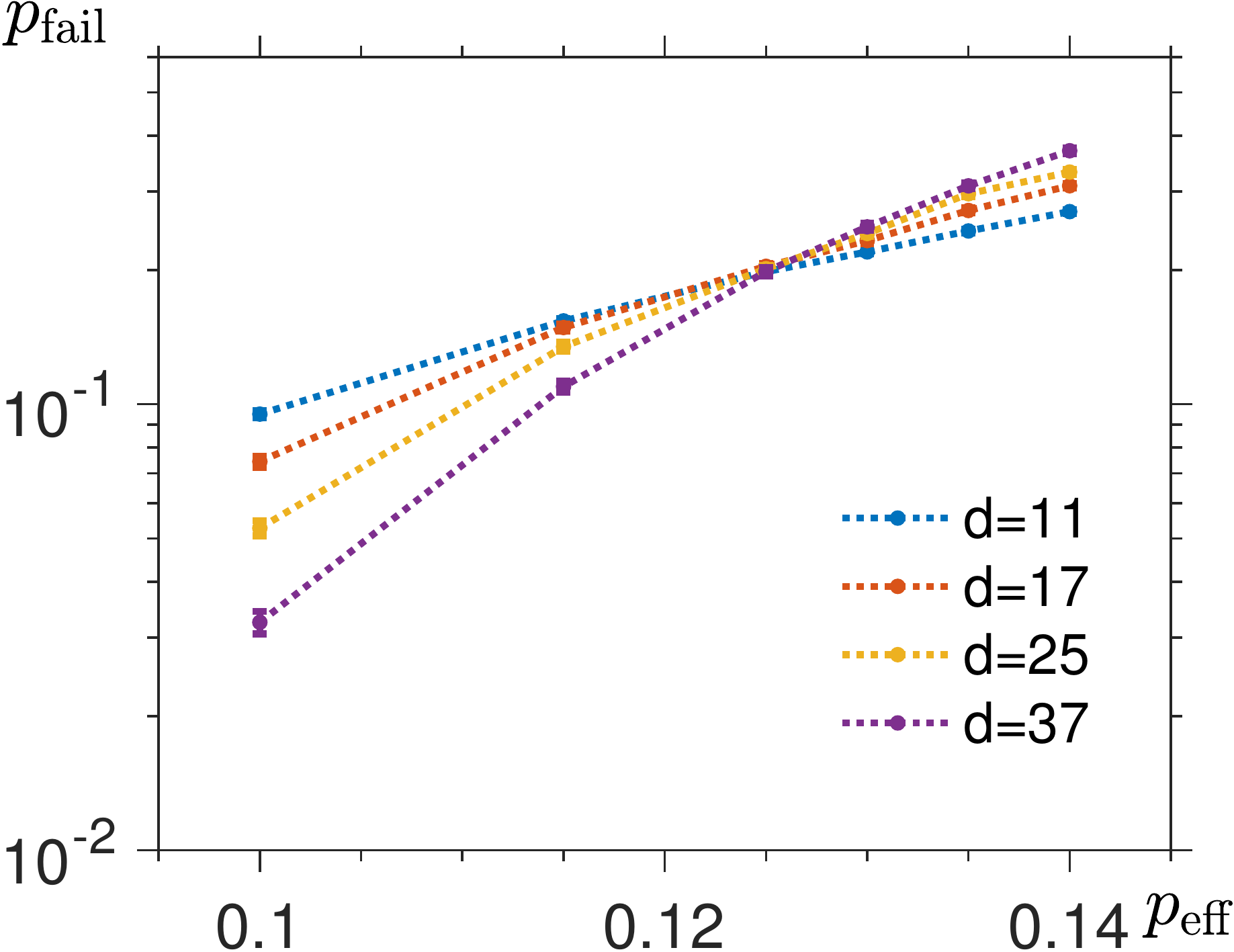}\quad
(f)\includegraphics[width= 0.29\textwidth]{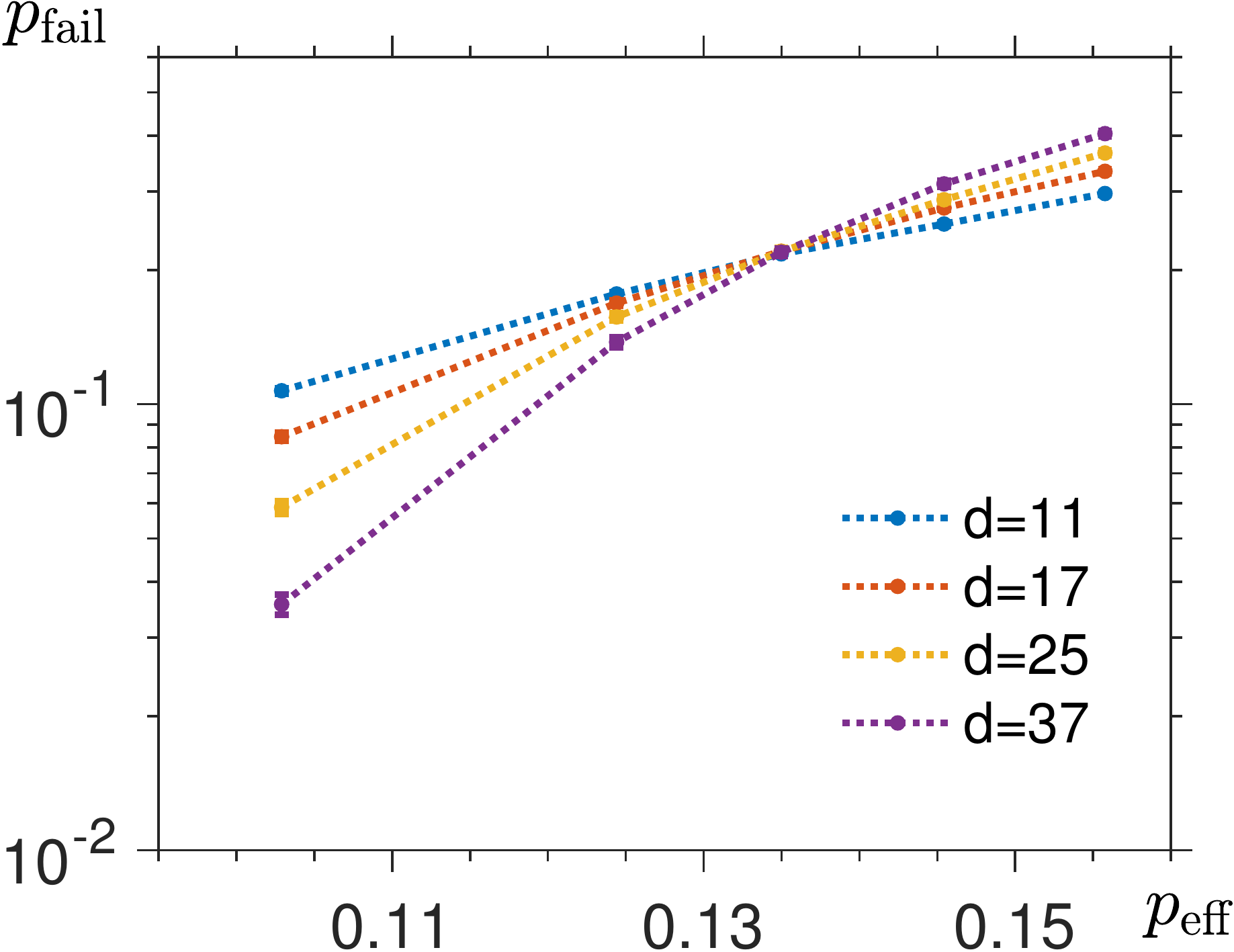}
\caption{
The failure probability $\pfail(\peff,d)$ of (a)-(c) the neural decoder and (d)-(f) the projection decoder for the 2D triangular color code of distance $d$ as a function of the effective error rate $\peff$.
We consider three noise models: (a),(d) bit-/phase-flip, (b),(e) depolarizing and (c),(f) NN-depolarizing.
We report that the neural decoder outperforms the projection decoder for all types of noise, exhibiting threshold near the optimal one.
}
\label{fig_thresholds_CC}
\end{figure*}

\begin{figure*}[ht!]
\centering
(a)\includegraphics[width= 0.29\textwidth]{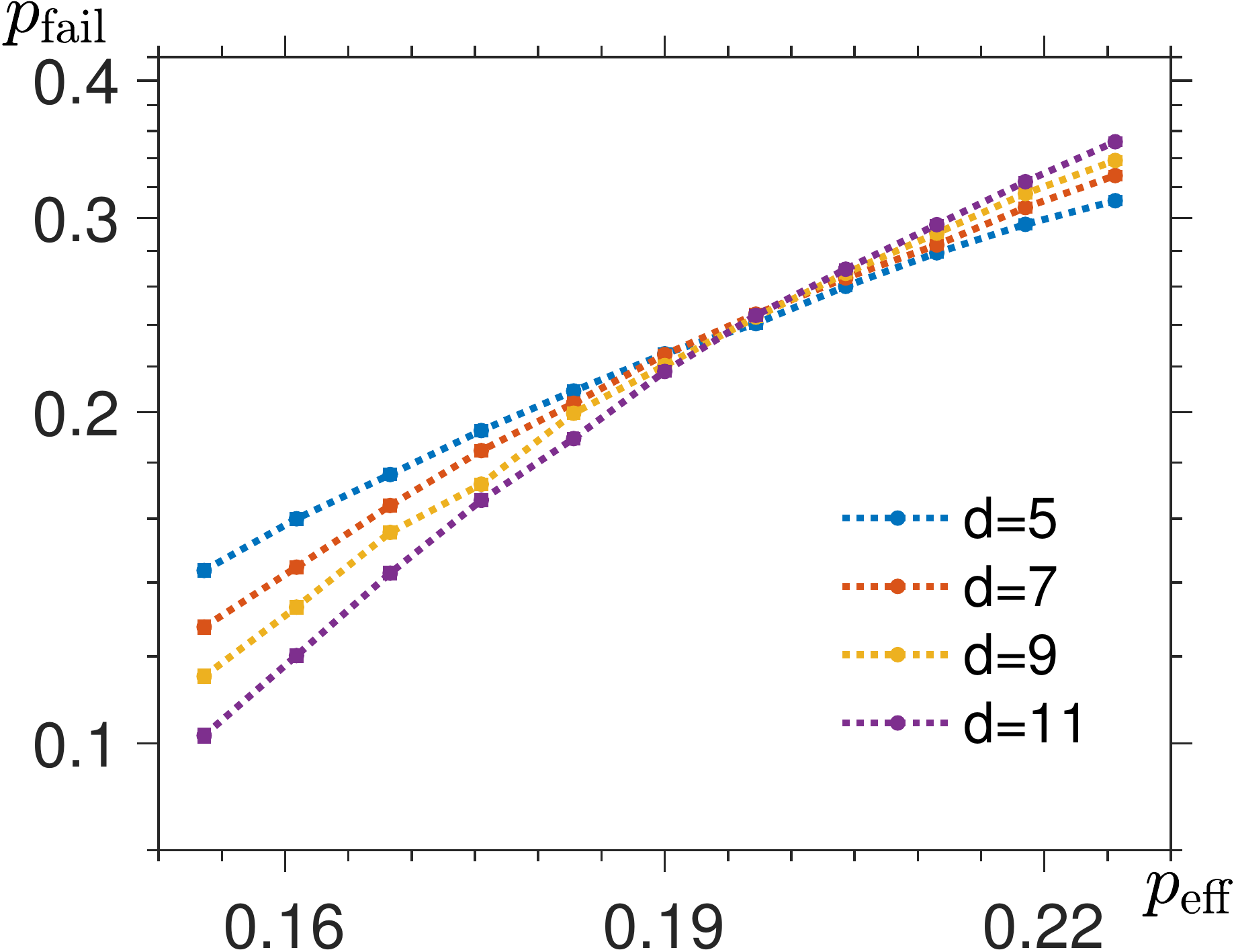}\quad
(b)\includegraphics[width= 0.29\textwidth]{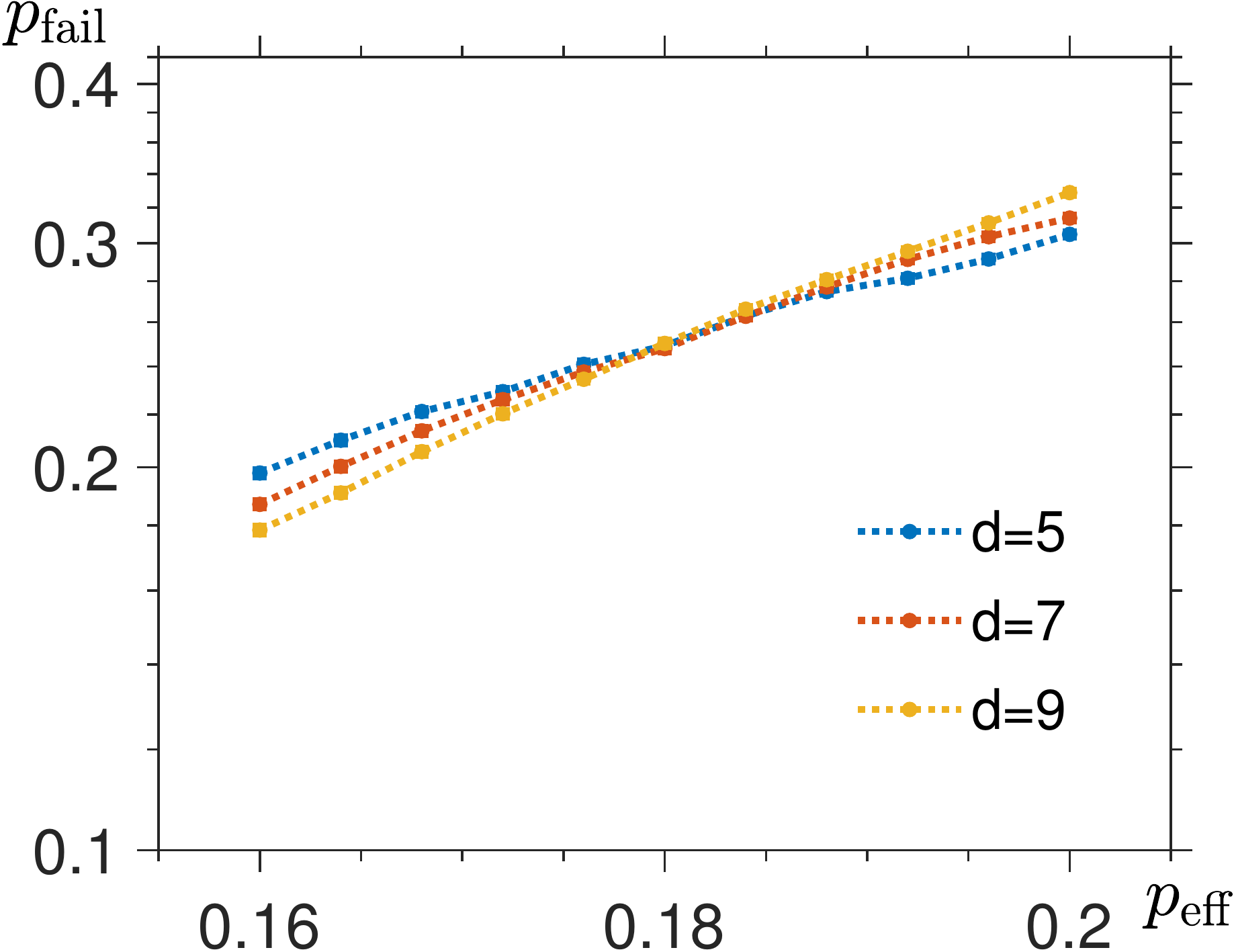}\quad
(c)\includegraphics[width= 0.29\textwidth]{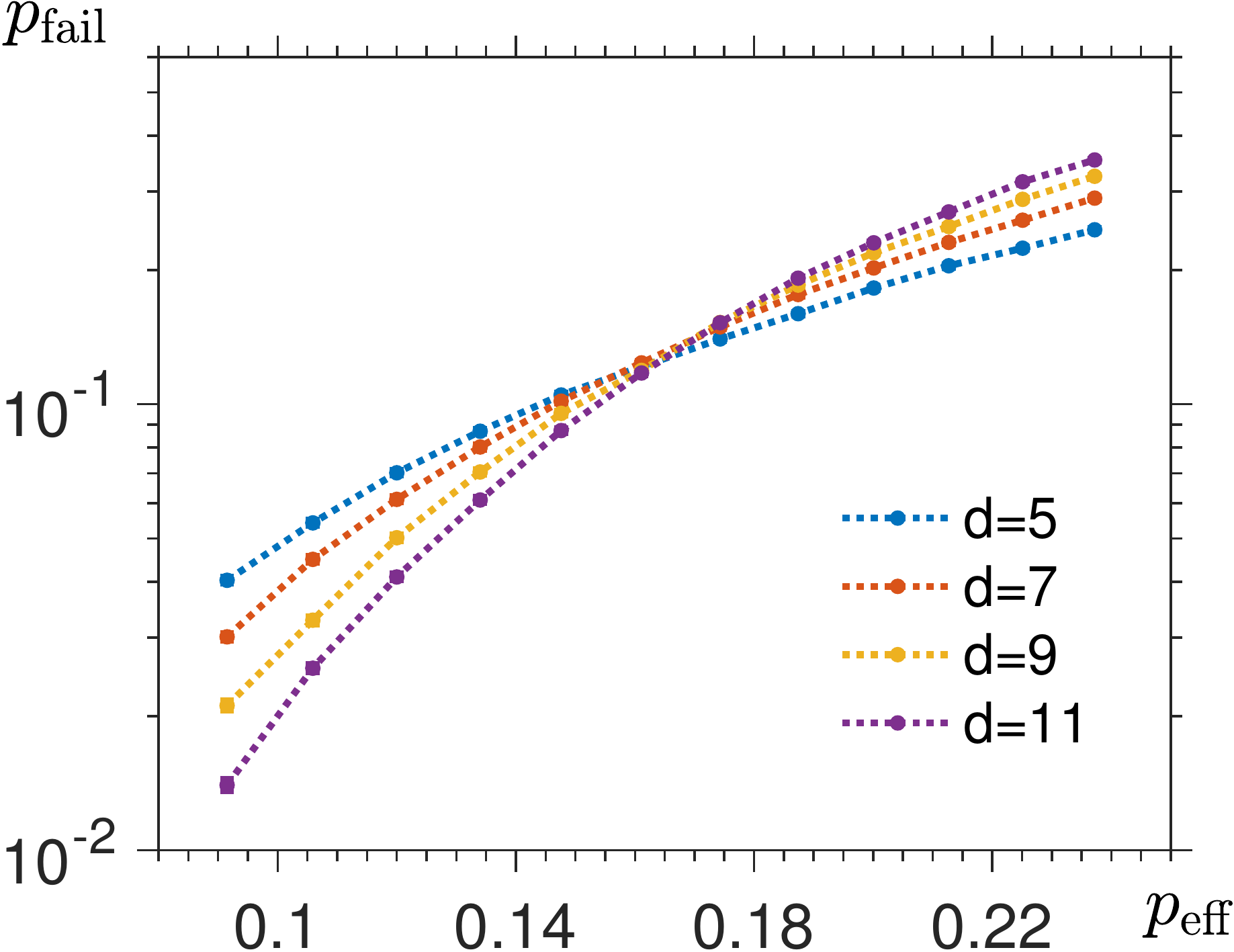}
(d)\includegraphics[width= 0.29\textwidth]{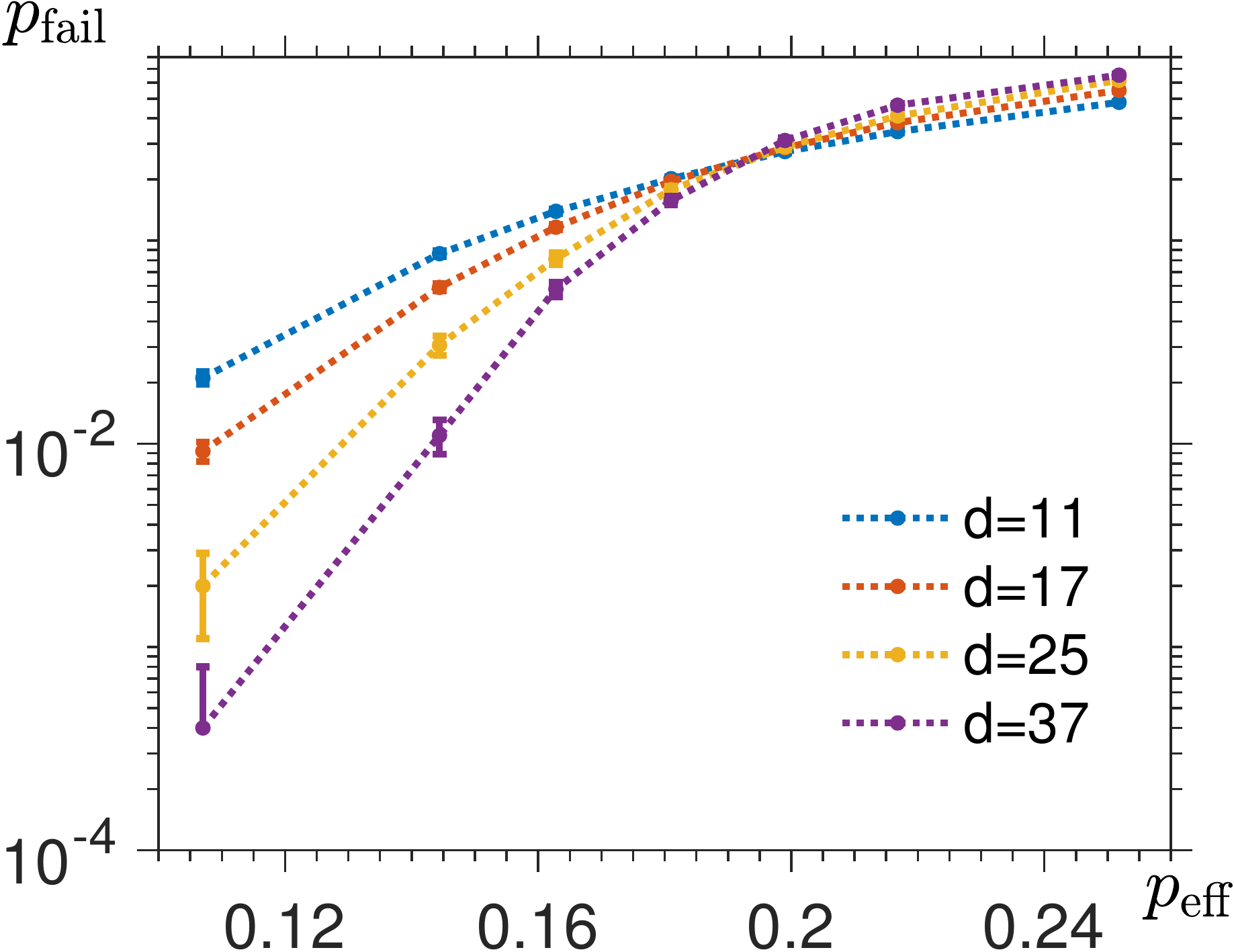}\quad
(e)\includegraphics[width= 0.29\textwidth]{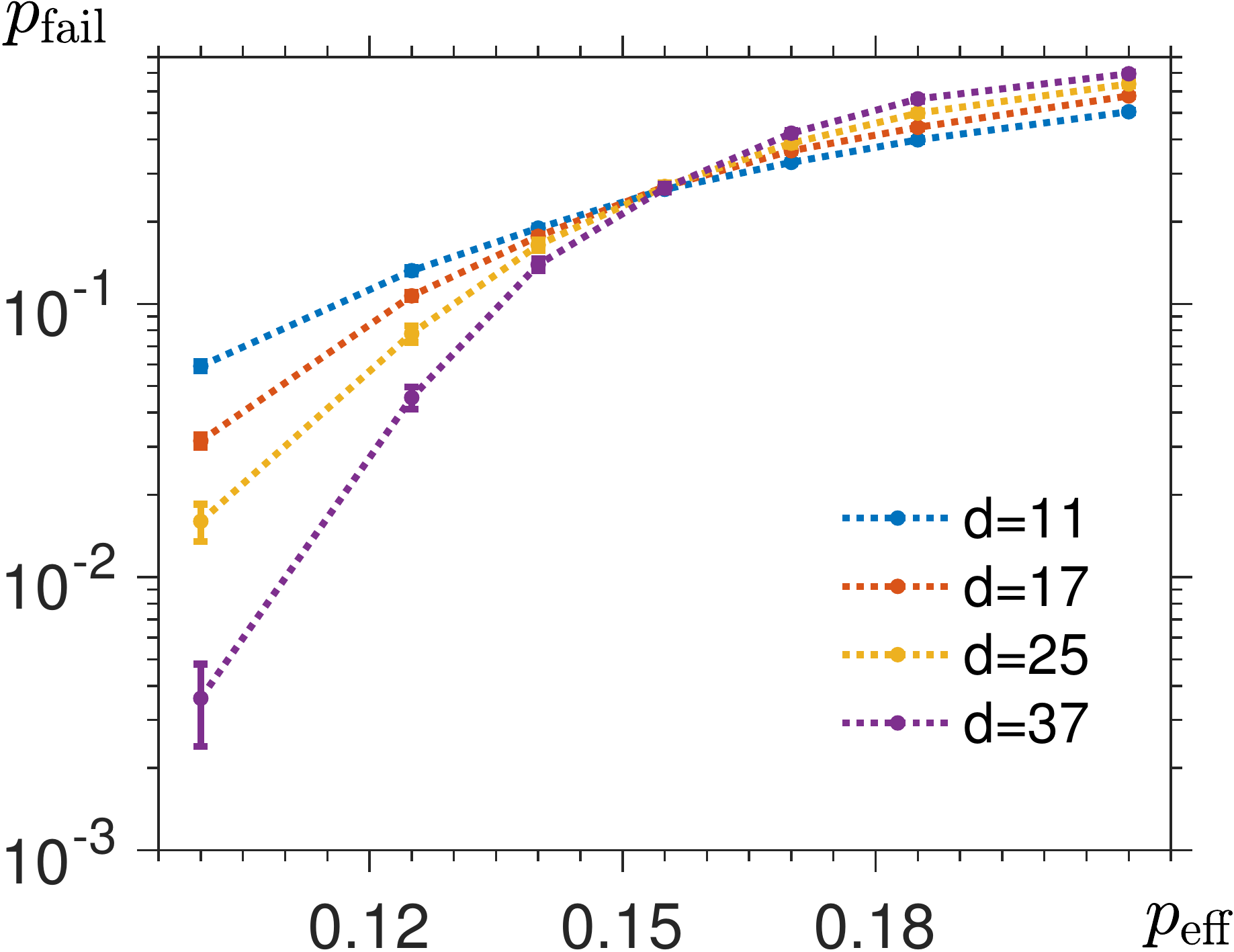}\quad
(f)\includegraphics[width= 0.29\textwidth]{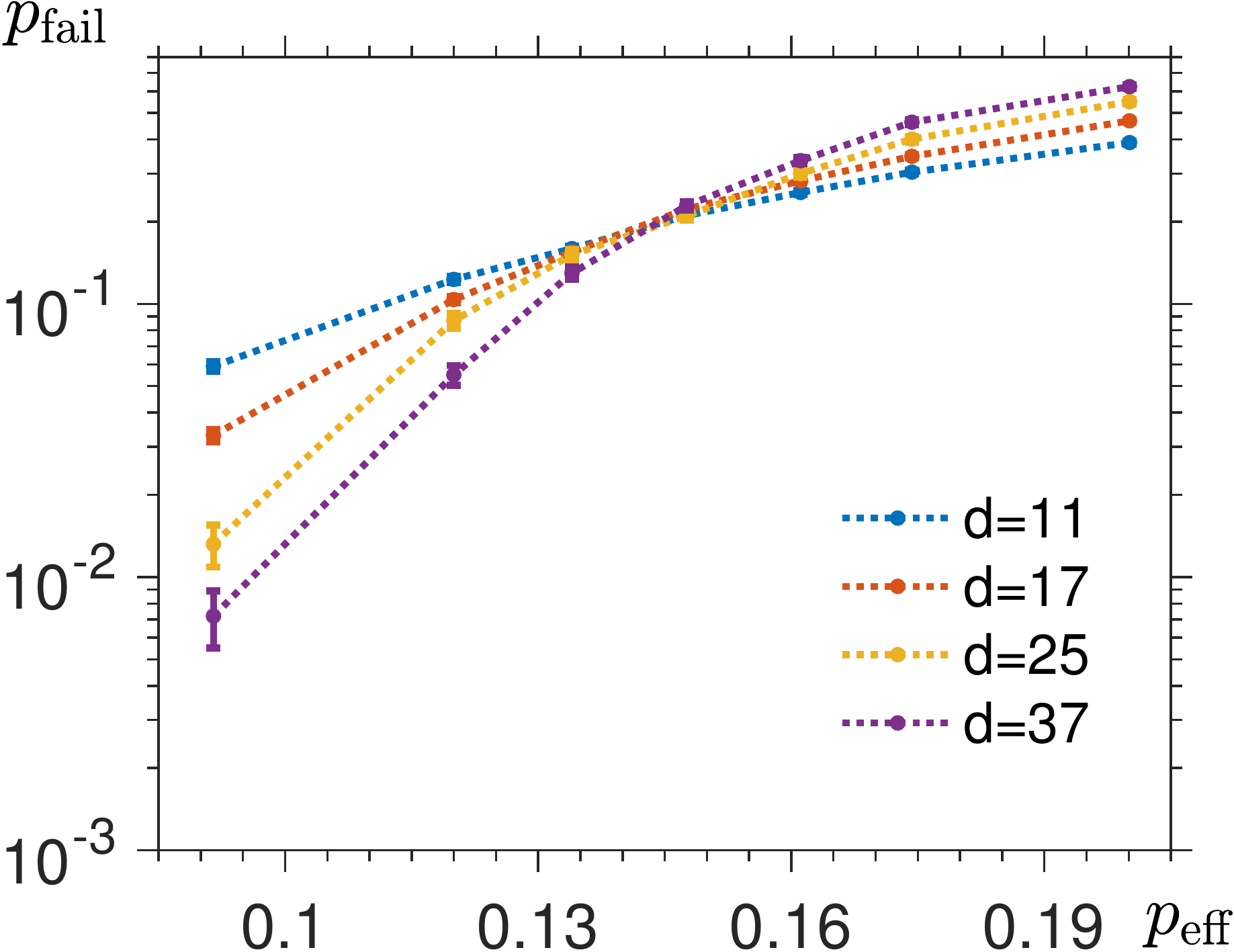}
\caption{
The failure probability $\pfail(\peff,d)$ of the (a)-(c) the neural decoder and (d)-(f) the Minimum-Weight Perfect Matching decoder for the 2D triangular toric code with a twist of distance $d$ as a function of the effective error rate $\peff$.
We consider three noise models: (a),(d) bit-/phase-flip, (b),(e) depolarizing and (c),(f) NN-depolarizing.
We report that the neural decoder significantly outperforms the Minimum-Weight Perfect Matching decoder for noise models with correlated errors and exhibits threshold near the optimal one.
}
\label{fig_thresholds_TC}
\end{figure*}

In order to assess the versatility of neural-network decoding, we qualitatively study its performance for the toric and color codes under three different noise models: bit-/phase-flip, depolarizing and NN-depolarizing.
First, we train a neural network for every code with the code distance up to $d=11$.
The optimized hyperparameters of considered neural networks are presented in Table~\ref{tab_hyperparameters}.
Then, we numerically find the decoder failure probability $\pfail(\peff,d)$ of the neural decoder as a function of the effective error rate $\peff$.
By plotting the decoder failure probability $\pfail(\peff,d)$ for different code distances $d$ and finding their intersection we numerically establish the existence of non-zero threshold for the neural decoder and estimate its value; see Figs.~\ref{fig_thresholds_CC}~and~\ref{fig_thresholds_TC}.

We benchmark the performance of the neural decoder against the leading efficient decoders of the toric and color code.
In particular, we analyze the standard decoders based on the Minimum-Weight Perfect Matching algorithm and the projection decoder.
In our implementation, we use the Blossom V algorithm provided by Kolmogorov \cite{Kolmogorov2009}.

We report that the neural decoder for the color code significantly outperforms the projection decoder for all considered noise models, even for the simplest bit-/phase-flip noise model.
The neural decoder threshold values we find approach the upper bounds from the maximal-likelihood decoder.
The neural decoder for the toric code shows comparable performance as the Minimum-Weight Perfect Matching decoder for the bit-/phase-flip noise, however offers noticeable improvements for correlated noise models. 
We remark that optimal decoding thresholds for topological codes can be found via statistical-mechanical mapping; see \cite{Dennis2002, Katzgraber2009, Bombin2012, Kubica2017}.
The threshold values we find are expressed in terms of the effective error rate $\peff$ and are listed in Table~\ref{tab_thresholds}.

As with all learning models, it is important to address the possibility of overfitting.
We know that the test samples are different (with high probability) from the training samples, since they are randomly chosen from a set that scales exponentially with the number of physical qubits.
We remark that the required training set seems to scale exponentially with the code distance, however it constitutes a vanishing fraction of all possible syndrome configurations.
Moreover, the classification accuracy on the test samples is the same as the final training accuracy.
Thus, we can conclude that the neural network learns to correctly label syndromes typical for the studied noise models, resulting in well-performing neural decoders.

\section{Discussions}

We have conclusively demonstrated that neural-network decoding for topological stabilizer codes is very versatile and clearly outperforms leading efficient decoders.
We focused on the triangular color code and the toric code a twist, whose physical qubits are arranged in the same way but their stabilizer groups are different.
We studied the performance of neural-network decoding for different noise models, including the spatially-correlated depolarizing noise.
In particular, we numerically established the existence of non-zero threshold and found significant improvements of the color code threshold over the previously reported values; see Table~\ref{tab_thresholds} and Figs.~\ref{fig_thresholds_CC}~and~\ref{fig_thresholds_TC}.
This result indicates that the relatively low threshold of the color code, which was considered to be one of its main drawbacks, can be easily increased, making quantum computation with the color code more appealing than initially perceived \cite{Wang2010, Fowler2011, Landahl2014}.

We emphasize that the neural network does not explicitly use any information about the topological code or the noise model.
The neural network is trained on very simple data usually available from the experiment, which includes the information about the measured syndrome and whether the simple deterministic decoding, i.e., the excitation removal algorithm, succeeds.
Importantly, this raw data can not only be used to train the neural network, but also to characterize the quantum device \cite{Combes2014}.
Without assuming any simplistic noise models the neural network efficiently detects the actual error patterns in the system and subsequently ``learns'' about the correlations between observed errors.
This provides a heuristic explanation why neural decoding is currently the best strategy to decode the color code, since the correlations between errors in the color code are difficult to account for in standard approaches \cite{Delfosse2014a}.
Using neural networks simplifies and speeds up the process of designing good decoders, which is rather challenging due to its heavy dependency on the choice of the quantum error-correcting code as well as the noise model.

Our results show that neural-network decoding can be successfully used for quantum error-correction protocols, especially in the systems affected by a priori unknown noise with correlated errors.
We stress that neural-network decoding already provides an enormous data-compression advantage over methods based on (partial) look-up tables, even for small-distance quantum codes.
However, an important question of scalability has to be addressed if neural decoders are ever going to be used for practical purposes on future fault-tolerant universal quantum devices.
One possible approach to scalable neural networks is to reduce the connectivity between the layers by exploiting the information about the topological code lattice and geometric locality of stabilizer generators.
We imagine incorporating convolutional neural networks as well as some renormalization ideas in the future scalable neural decoders.
Also, a fully-fledged neural decoder should account for the possibility of faulty stabilizer measurements \cite{chamberland2018}.
We do not perceive any fundamental reasons why neural-network decoding, possibly based on recurrent neural networks, would not work for the circuit level noise model.
However, in that setting the training dataset as well as the size of the required neural network grow substantially, making the training process computationally very challenging.

\begin{acknowledgments}
We would like to thank Ben Brown, Jenia Mozgunov and John Preskill for valuable discussions, as well as Evert van Nieuwenburg for his feedback on this manuscript.
During the preparation of the manuscript two related preprints were made available \cite{Davaasuren2018, Jia2018}, however their scope and emphasis are different from our work.
NM acknowledges funding provided by the Caltech SURF program.
AK acknowledges funding provided by the Simons Foundation through the ``It from Qubit'' Collaboration. Research at Perimeter Institute is supported by the Government of Canada through Industry Canada and by the Province of Ontario through the Ministry of Research and Innovation.
TJ acknowledges the support from the Walter Burke Institute for Theoretical Physics in the form of the Sherman Fairchild Fellowship. The authors acknowledge the support from the Institute for Quantum Information and Matter~(IQIM).
\end{acknowledgments}

\bibliographystyle{apsrev4-1}
\bibliography{biblio_NNdecoder}

\end{document}